\documentclass[trackchanges, twocolumn]{aastex701}
\usepackage[utf8]{inputenc}
\usepackage{amsmath,amssymb,graphicx,booktabs,bm}
\usepackage{nccmath}
\usepackage{tabularx}
\usepackage{comment}
\usepackage{float}
\usepackage{xcolor}
\usepackage{soul}
\usepackage{comment}

\setstcolor{red}  

\shorttitle{Probing reionization with $\mathrm{kSZ}^{2}\times21\,\mathrm{cm}^{2}$ cross-correlation}
\shortauthors{Guo et al.}

\begin{document}

\title{Probing Early Phases of the Epoch of Reionization with the $\mathrm{kSZ}^{2}\times21\,\mathrm{cm}^{2}$ Cross-Correlation}
\author{Wen-Qing Guo}
\affiliation{Department of Physics, Stellenbosch University, Matieland 7602, South Africa}
\email{}
\author{Yin-Zhe Ma}
\altaffiliation{Corresponding author: \url{mayinzhe@sun.ac.za}; South Africa Research Chair (SARChI)}
\affiliation{Department of Physics, Stellenbosch University, Matieland 7602, South Africa}
\email{}
\author{Adam Lidz}
\affiliation{Center for Particle Cosmology, Department of Physics and Astronomy, University of Pennsylvania, Philadelphia, Pennsylvania 19104, USA}
\email{}
\author{Steven G. Murray}
\affiliation{Department of Physics, Stellenbosch University, Matieland 7602, South Africa}
\email{}
\author{Hemanth Potluri}
\affiliation{Department of Physics, Stellenbosch University, Matieland 7602, South Africa}
\affiliation{Kapteyn Astronomical Institute, University of Groningen, PO Box 800, NL-9700 AV Groningen, the Netherlands}
\email{}

\begin{abstract}
Observations of the Epoch of Reionization (EoR) provide key insights into the nature of the first luminous sources and the thermal state of the intergalactic medium (IGM). Cross-correlations between the kinetic Sunyaev--Zel'dovich (kSZ) effect and neutral hydrogen (HI) 21-cm emission offer a promising probe of this era, sensitive to both the ionization and thermal histories of the early Universe. We investigate the cross-correlation of the squared kSZ field with the squared 21-cm intensity-mapping field ($\mathrm{kSZ}^{2}\times21\mathrm{cm}^{2}$) over $z\simeq9$--$16$, focusing on the impact of spatial fluctuations in the HI spin temperature. This higher-order statistic avoids line-of-sight cancellations in the kSZ signal, while squaring the 21-cm field enables non-vanishing correlations with pairs of equal and opposite high-$k_\parallel$ modes accessible outside the foreground wedge. We forecast the cumulative signal-to-noise ratio ($\mathrm{S/N}$) for several combinations of 21-cm and CMB experiments, including realistic instrumental noise. For a fiducial heating history and an optimistic foreground-avoidance scenario ($k_\parallel\leq0.1\,h\,\mathrm{Mpc}^{-1}$), we find $\mathrm{S/N}\simeq5.89$ for SKA-Low AA$^\ast\times$CMB-HD and $\simeq3.62$ for HERA$\times$CMB-HD. If the foreground wedge extends to the horizon, these decrease to $2.94$ and $2.85$, respectively.  {Our results demonstrate that $\mathrm{kSZ}^{2}\times21\mathrm{cm}^{2}$ is sensitive to the ionization structure of the IGM, and, provided that X-ray heating is not too efficient, also to its thermal evolution, providing complementary constraints on the high-redshift thermal and reionization 
histories.}

\end{abstract}

\section{Introduction}\label{sec:intro}
The Epoch of Reionization (EoR) represents one of the most fundamental phase transitions in cosmic history, during which the predominantly neutral intergalactic medium (IGM) was progressively heated and ionized by X-ray and ultraviolet radiation produced by the first generations of stars, galaxies, and accreting black holes \citep{Barkana:2000fd, Benson:2005cc, Zaroubi2013}. A comprehensive characterization of the EoR is essential for developing a complete picture of early structure formation and the evolution of the Universe. Despite important progress, however, the timing, topology, and astrophysical drivers of reionization remain key open questions in modern cosmology \citep{Fan:2006dp,Dai2019}.

A broad range of observational probes has been developed to constrain the physical processes associated with reionization. High-redshift quasar absorption spectra provide important constraints on the late stages of reionization through the Gunn--Peterson trough \citep{Miralda-Escude:1997ham, Becker2001, Fan2002}, as well as Ly$\alpha$/Ly$\beta$ forest dark-pixel and damping-wing analyses \citep{McGreer:2014qwa, Greig:2016vpu, Davies:2018yfp, Greig:2018rts, Wang:2020zae}, although these measurements probe only a limited number of one-dimensional lines of sight. Measurements of the Thomson-scattering optical depth $\tau$ from the Cosmic Microwave Background (CMB) provide complementary, integrated constraints on the global reionization history~\citep{Planck2016_reionization,Qin:2020xrg, Pagano:2019tci}. Observations of high-redshift galaxies and Lyman-$\alpha$ emitters further constrain the IGM neutral fraction through the attenuation and transmission of Lyman-$\alpha$ photons \citep{Sobacchi:2015gpa, Mason:2019ixe, Whitler:2019nul, Weinberger:2018hvi, Pagano:2021rye, Prieto-Lyon2026}, but remain subject to uncertainties from radiative transfer and line-of-sight attenuation, as well as limited sensitivity at $z\gtrsim7$.

Among the various proposed probes, the 21-cm signal from the hyperfine transition of neutral hydrogen (HI) is widely regarded as one of the most promising direct tracers of the three-dimensional (3D) structure of the IGM during the EoR \citep{Furlanetto:2006jb, Morales2010, Pritchard:2011xb}. Dedicated radio interferometers, including the Low-Frequency Array (LOFAR; \citealt{Beardsley:2016njr, Patil:2017zqk, Ghara:2020syx}), the Murchison Widefield Array (MWA; \citealt{Lidz:2007az, Trott:2020szf, Greig:2020suk, Yoshiura:2021yfx}), the Hydrogen Epoch of Reionization Array (HERA; \citealt{HERA:2021bsv, HERA:2021noe, HERA:2022wmy, HERA:2025ajm}), and the forthcoming Square Kilometre Array (SKA; \citealt{Mellema:2012ht, Weltman:2018zrl, EorCDScienceWorkingGroup:2026trp}), aim to detect the 21-cm signal and ultimately characterize its statistical properties, most notably through the 21-cm power spectrum. However, direct measurements of the 21-cm auto-power spectrum remain extremely challenging because the cosmological signal is contaminated by astrophysical foregrounds that can exceed it by several orders of magnitude \citep{Ghosh2012, Liu:2014bba, Liu:2014yxa, Thyagarajan:2015ewa, Bull:2026pbi}, in addition to ionospheric effects, calibration uncertainties, and complex instrumental systematics. Cross-correlating 21-cm observations with independent cosmological tracers therefore provides a powerful complementary strategy, as uncorrelated foregrounds and instrumental systematics are expected to be strongly suppressed in the cross-correlation \citep{Ma:2017gey, LaPlante:2022nlp, Moriwaki:2024kvp, Fronenberg:2024olu, Chen:2025hzj, Gagnon-Hartman:2025oxd, Zhou:2025fgv}. In this context, galaxy surveys and metal line-intensity mapping are promising complementary probes in their own right, and
can be cross-correlated with 21-cm data. Their reach, however, is
limited to the later stages of reionization. 
Accessing the earlier stages of reionization, in which spatial fluctuations of the 21cm spin temperature play a larger role, is considerably more challenging with
these tracers. For this early phase, cross-correlations between 21-cm
observations and CMB measurements therefore remain the only current
prospects.

Among CMB probes, a particularly promising candidate is the kinetic Sunyaev--Zel'dovich (kSZ) effect, which arises from the Doppler shifting of CMB photons by Thomson scattering off free electrons with non-zero peculiar velocities \citep{Birkinshaw:1998qp,Ma:2013taq}. Because the kSZ signal traces the momentum field of free electrons, it contains valuable information about the ionization structure of the high-redshift Universe and is therefore naturally connected to the distribution of neutral hydrogen during reionization. However, a simple linear cross-correlation, $\mathrm{kSZ} \times \mathrm{21\,cm} $, suffers from cancellations because positive and negative kSZ signals occur with approximately equal likelihood, reflecting the random signs of the line-of-sight peculiar velocities \citep{Dore:2003ex}. A common approach to overcoming this cancellation is to consider the squared kSZ field, which removes the sign information of the kSZ fluctuations and allows the magnitude of the kSZ signal to be correlated with the HI field through higher-order statistics \citep{Ferraro:2016ymw, Ma:2017gey}.

Nevertheless, the bispectrum cross-correlation $\rm  \mathrm{kSZ}^2 \times {\rm 21\,cm} $ encounters a critical observational barrier, since the kSZ signal is a line-of-sight integrated two-dimensional (2D) map, its parallel wavevector component is intrinsically zero ($k^\parallel_{\mathrm{kSZ}} = 0$). Under the momentum conservation condition $k^\parallel_{\rm HI} + k^\parallel_{\mathrm{kSZ},1} + k^\parallel_{\mathrm{kSZ},2} = 0$, this enforces $k^\parallel_{\rm HI} = 0$ for the HI 21-cm field. Since astrophysical foregrounds are spectrally smooth, they concentrate heavily at low parallel wavevectors ($k^\parallel \approx 0$), making the $k^\parallel_{\rm HI} = 0$ modes completely unobservable inside the so-called 21-cm \textit{foreground wedge} \citep{Liu:2014bba, Liu:2014yxa}. To circumvent this limitation, one must square the 21-cm field as well, evaluating the trispectrum cross-correlation $\rm \mathrm{kSZ}^2 \times {\rm 21\,cm}^2$. This configuration relaxes the parallel wavevector constraint to $k^\parallel_{\rm HI,1} + k^\parallel_{\rm HI,2} = 0$, allowing non-zero $k^\parallel$ modes to be probed while safely escaping the foreground-contaminated wedge region.

\begin{figure}
    \centering
    \includegraphics[width=1.\linewidth]{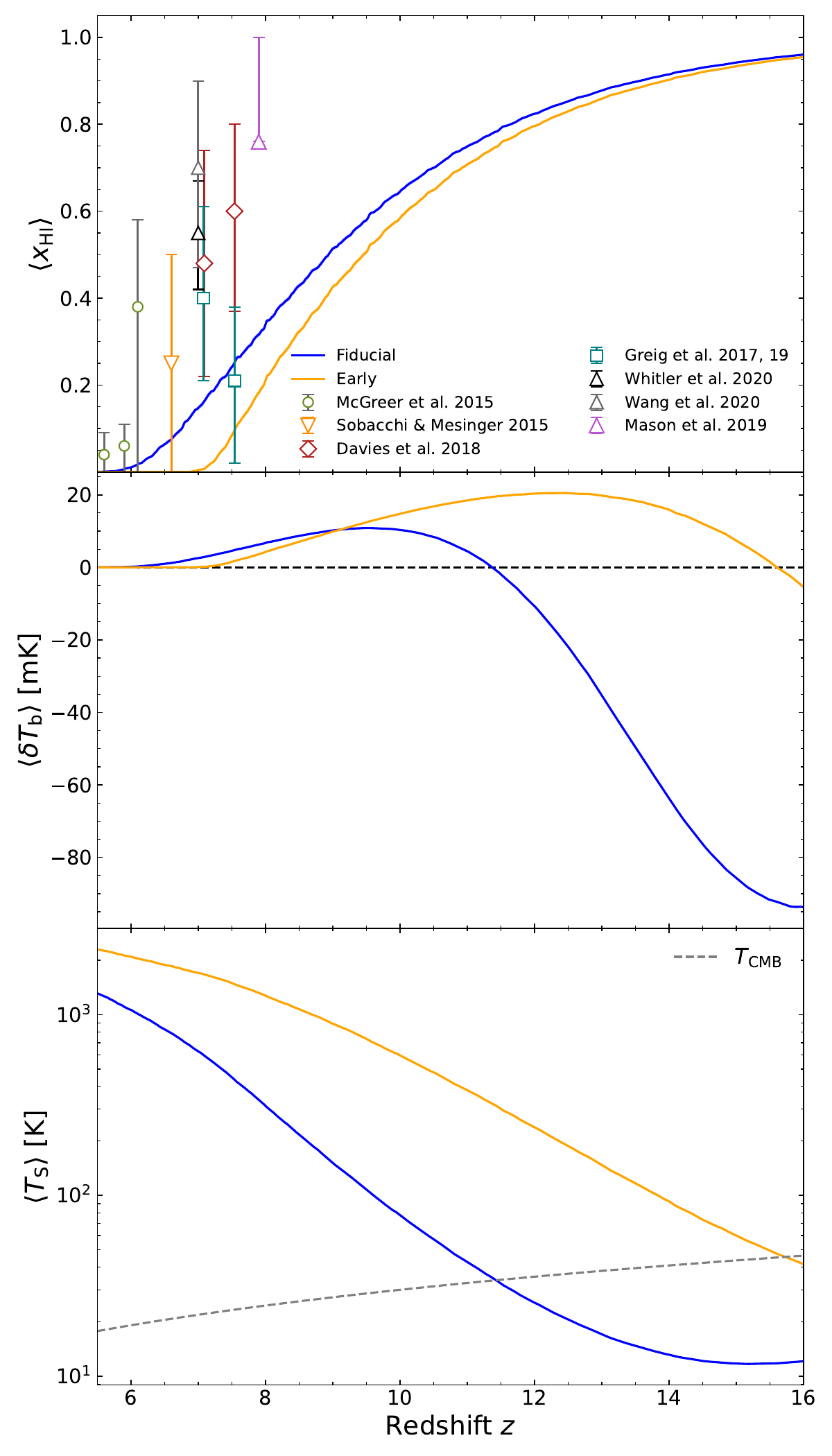}
    \caption{ {Evolution of the volume-averaged neutral hydrogen fraction, 21-cm differential brightness temperature, and HI spin temperature for the fiducial (blue curves) and early-heating (orange curves) scenarios. \textit{Upper:} Volume-averaged neutral hydrogen fraction as a function of redshift, together with observational constraints (see also, e.g. Table in~\citealt{Dai2019}). \textit{Middle:} Corresponding evolution of the volume-averaged 21-cm differential brightness temperature. \textit{Lower:} Evolution of the volume-averaged HI spin temperature $\langle T_{\rm s} \rangle$ as a function of redshift. The dashed grey line denotes the CMB temperature $T_{\rm CMB}(z)$ as a reference.}
   }
    \label{fig:NeutralFraction_BrightnessTemperature}
\end{figure}

While early pioneering work by \cite{Zhou:2025fgv} demonstrated the feasibility of the $\rm  \mathrm{kSZ}^2 \times {\rm 21\,cm}^2 $ cross-correlation and presented forecasts for its signal-to-noise ratio (S/N) over the redshift range $z\sim7$--$11$, key physical and observational gaps remain. First, previous studies relied on simulations (e.g., \texttt{zreion}~\citep{Battaglia2013}) that adopt the saturated heating assumption ($T_\mathrm{s} \gg T_{\mathrm{CMB}}$), explicitly neglecting HI spin temperature ($T_\mathrm{s}$) spatial fluctuations. 
Although this approximation may be appropriate during the later stages of reionization, it breaks down at earlier epochs, when the thermal state of the IGM is still evolving and X-ray heating remains highly inhomogeneous. During this stage, spatial fluctuations in $T_{\mathrm{s}}$ can strongly modulate the 21-cm differential brightness temperature and modify its non-Gaussian statistical properties. As a consequence, the amplitude and scale dependence of the $\mathrm{kSZ}^2 \times {\rm 21\,cm}^2$ cross-power spectrum may be significantly affected.

Second, previous detectability forecasts have generally adopted simplified instrumental noise prescriptions and therefore do not fully capture several observational effects relevant to realistic interferometric measurements. In particular, the sensitivity of an interferometric array depends on its baseline distribution, frequency-dependent $uv$ coverage, and the synthesis of Fourier modes through Earth rotation. These effects can substantially modify the instrumental response and sensitivity across angular scales and should therefore be incorporated when assessing the detectability of the $\mathrm{kSZ}^2 \times {\rm 21\,cm}^2$ signal with current and future radio facilities.

In this work, we address these limitations through a comprehensive investigation of the $\mathrm{kSZ}^2 \times {\rm 21\,cm}^2$ cross angular power spectrum over the high-redshift range $z=9$--$16$, using angular lightcones generated with the semi-numeric simulation \texttt{21cmFASTv4}\footnote{\url{https://21cmfast.readthedocs.io/en/stable/}} \citep{Mesinger:2010ne, Murray:2020trn, Davies:2025wsa}. Our analysis has two primary objectives. First, we quantify the impact of spatial fluctuations in the HI spin temperature on the $\mathrm{kSZ}^2 \times {\rm 21\,cm}^2$ cross-correlation during the early stages of reionization. Second, we perform realistic forecasts for the detectability of the cross-correlation signal by combining HI observations from HERA {-350 \citep{Berkhout_2024} (hereafter HERA)} and SKA-Low AA$^{*}$  {(comprising 307 stations \footnote{\url{https://www.skao.int/sites/default/files/documents/SKAO-TEL-0002299_01_staged_delivery.pdf}})} with kSZ measurements from the Simons Observatory \citep{SimonsObservatory:2018koc} and CMB-HD \citep{Sehgal:2019ewc}. For the HI observations, we incorporate Earth-rotation synthesis and realistic $uv$ sampling through the \texttt{21cmSense} framework \footnote{\url{https://21cmsense.readthedocs.io/en/latest/}} \citep{Murray:2024the}. By combining these two effects, this work provides an important extension of, and complement to, the $\mathrm{kSZ}^2 \times {\rm 21\,cm}^2$ cross-correlation framework developed by \cite{Zhou:2025fgv}, particularly towards the earlier stages of the EoR.

\begin{figure*}[t]
    \centering
    \includegraphics[width=0.48\linewidth]{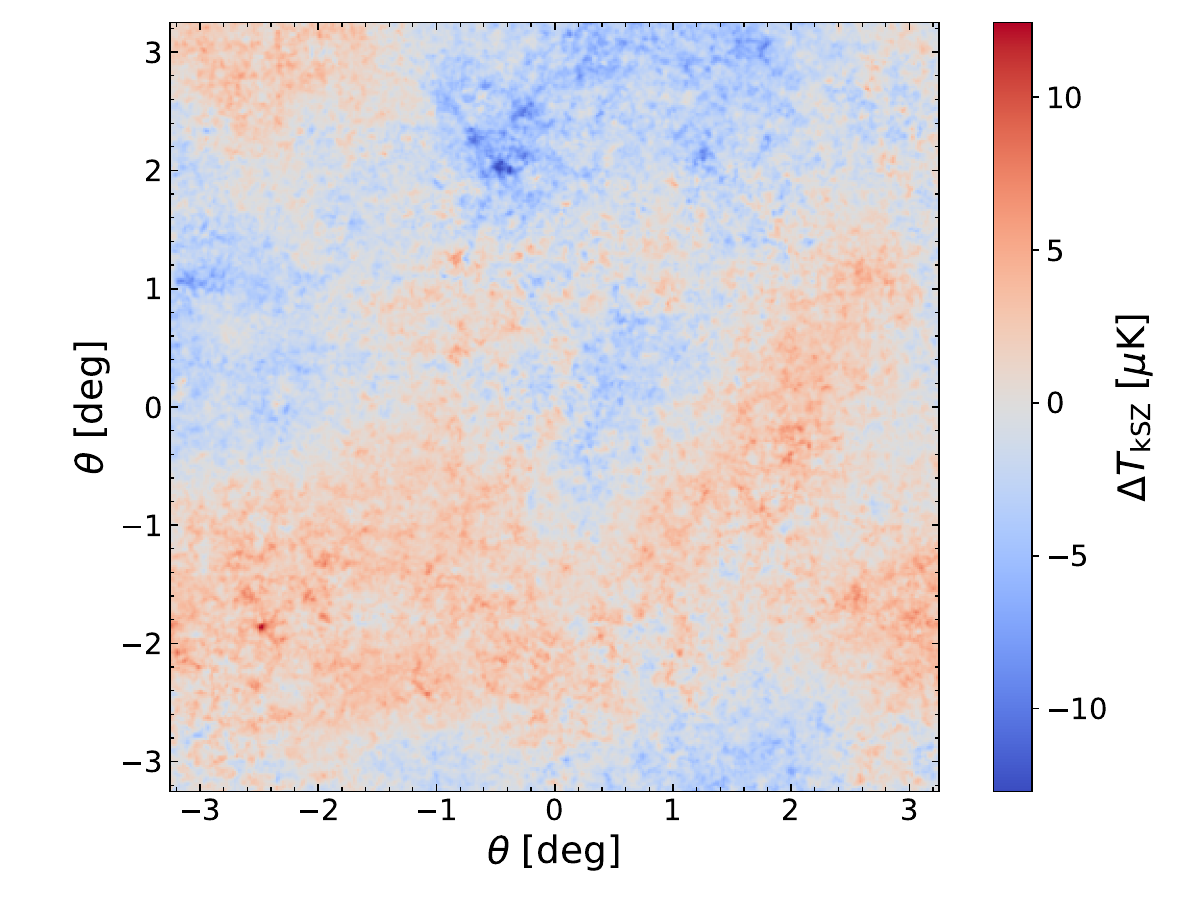}
    \includegraphics[width=0.48\linewidth]{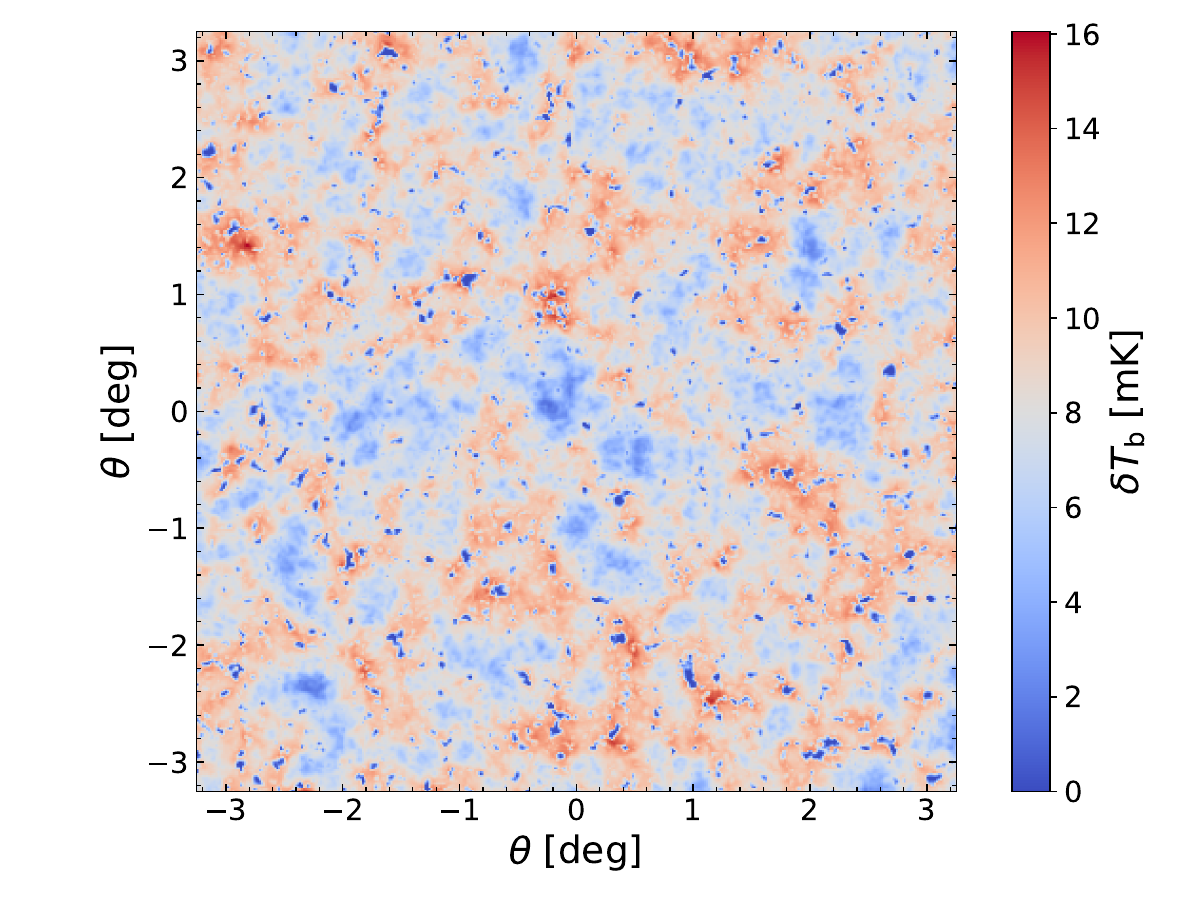}
    \caption{Representative simulated maps from the fiducial model. \textit{Left}--High-$z$ EoR kSZ signal integrated over $z=9$--$16$. \textit{Right}--A slice of the 21-cm differential brightness temperature field at $z=10$.}
    \label{fig:kSZ_HI_maps}
\end{figure*}

The remainder of this paper is organized as follows. In Sec.~\ref{sec:simulations}, we describe the semi-numerical simulations used to model the reionization history and to generate the kSZ and 21-cm signals. In Sec.~\ref{sec:CAPS}, we present the simulated $\mathrm{kSZ}^2 \times {\rm 21\,cm}^2$ angular cross-power spectrum and investigate its dependence on the spin temperature, filtering procedure, and heating history. In Sec.~\ref{sec:detectability}, we present forecasts for the detectability of the signal. Finally, we conclude in Sec.~\ref{sec:conclusion}. Throughout this paper, we assume a spatially flat $\Lambda$CDM cosmology with cosmological parameters consistent with the \textit{Planck} 2018 results: $H_{0}=67.32~{\rm km\,s^{-1}Mpc^{-1}}$, $\Omega_{\rm c}h^{2}=0.1201$, and $\Omega_{\rm b}h^{2}=0.02238$ \citep{Planck:2018vyg}.

\section{Simulations}\label{sec:simulations}

\subsection{HI 21-cm Field}

The cosmological 21-cm signal is characterized by the differential brightness temperature relative to the CMB, which depends on the local density, neutral hydrogen fraction, and thermal state of the IGM. In the optically thin limit, the differential brightness temperature at redshift $z$ and sky direction $\hat{r}$ can be approximated as \citep{Furlanetto:2006jb}
\begin{eqnarray}
\label{eq:HI}  
\delta T_{\rm b}(z,\hat{r}) &\simeq & \, 26\,x_{\rm HI}(z, \hat{r})\left[1+\delta_{\rm b}(z,\hat{r})\right] \nonumber \\ & \times & \left[1-\frac{T_{\rm CMB}(z)}{T_{\rm s}(z, \hat{r})}\right] \left[\frac{1 - Y_{\rm p}}{1 - 0.248}\right]\left(\frac{\Omega_{\rm b}}{0.044}\right) \nonumber \\ 
&\times & \left[\left(\frac{0.27}{\Omega_{\rm m}}\right)\left(\frac{1+z}{10}\right)\right]^{1/2}{\rm mK}
\end{eqnarray}
where $x_{\rm HI}$ denotes the neutral hydrogen fraction, $\delta_{\rm b}$ is the baryonic overdensity, $T_{\rm s}$ is the HI spin temperature, $T_{\rm CMB}$ is the CMB temperature, and $Y_{\rm p}$ is the primordial helium mass fraction. The factor $1-T_{\rm CMB}/T_{\rm s}$ determines whether the 21-cm signal is observed in absorption or emission. Consequently, spatial fluctuations in the ionization, density, and spin-temperature fields all contribute to the spatial structure and statistical properties of the 21-cm signal.

\subsection{Kinetic Sunyaev--Zel'dovich Field}

The kSZ effect arises from the Doppler shift of CMB photons that are Thomson-scattered by free electrons with non-zero peculiar velocities. The resulting CMB temperature fluctuation along a sky direction $\hat{r}$ is given by \citep{Sunyaev1980,CHM2015,Alvarez:2015xzu,Ferraro:2016ymw,Ma2017,Li2018,Planck2018_dispersion}
\begin{equation}\label{eq:kSZ_1} 
\begin{aligned} 
\frac{\Delta T_{\rm kSZ}}{T_{\rm CMB}}(\hat{r}) &=-\int_{0}^{\infty}\frac{\sigma_{\rm T}\,{\rm d}z}{(1+z)H(z)}\,n_{\rm e}(z,\hat{r})\,e^{-\tau}\left(\vec{v}\cdot \hat{r}\right), \end{aligned} 
\end{equation}
where $\sigma_{\rm T}$ is the Thomson scattering cross-section and $\vec{v}$ is the peculiar velocity of the electrons. We adopt a flat $\Lambda$CDM Hubble parameter $H(z) = H_{0}\sqrt{\Omega_{\rm m}(1+z)^{3} + \Omega_{\Lambda}}$, where $\Omega_{\Lambda}=1-\Omega_{\rm m}$ and $\Omega_{\rm m}=0.3158$. Assuming that helium is predominantly singly ionized during the EoR, and approximating the ionization fraction of helium by that of hydrogen, $x_{\rm HeII}\simeq x_{\rm HII}$, the physical number density of free electrons from both hydrogen and helium can be written as
\begin{eqnarray}\label{eq:ne}
    n_{\rm e} &=& \frac{\Omega_{\rm b}\rho_{\rm cr}}{ m_{\rm p}}(1+z)^{3}\left(1-\frac{3}{4}Y_{\rm p}\right) \nonumber \\
    & \times & \left[1+\delta_{\rm b }(z,\hat{r})\right]\left[ 1- x_{\rm HI}(z, \hat{r})\right],
\end{eqnarray}
where $\rho_{\rm cr}=3H_{0}^{2}/8\pi G=1.879\,h^{2}\times10^{-29}\,{\rm g\,cm^{-3}}$, and $m_{\rm p}$ is the proton mass. The term $\tau(z)$ in Eq. (\ref{eq:kSZ_1}) is the Thomson optical depth between the observer and redshift $z$, which is given by 
\begin{equation}\label{eq:tau}
    \tau(z) = \sigma_{\rm T}\int_{0}^{z}\frac{c\,{\rm d}z'}{(1+z')H(z')}n_{\rm e}(z').
\end{equation}
Here, we compute $e^{-\tau (z)}$ using the global average reionization history, since the Thomson optical depth is small and its line-of-sight fluctuations have a negligible impact on the kSZ signal \citep{LaPlante:2021ced}.

\subsection{Lightcones}\label{subsec:lightcones}
We generate realizations of the lightcones using the publicly available semi-numerical simulation code \texttt{21cmFASTv4}. This code combines perturbation theory with the excursion-set formalism to efficiently model the evolution of the matter density, neutral hydrogen fraction, HI spin temperature, and peculiar velocity fields during the cosmic dawn and the EoR. The density, neutral hydrogen fraction, and spin-temperature fields are subsequently used to construct the 21-cm differential brightness temperature field, while the density, neutral hydrogen fraction, and peculiar-velocity fields are used to compute the kSZ signal.

Our primary analysis focuses on the cross-correlation during the epoch in which spatial fluctuations in the HI spin temperature play an important role. To obtain sufficiently robust statistics for the $\mathrm{kSZ}^2 \times {\rm 21\,cm}^2$ cross-correlation, we generate 50 independent angular lightcones covering the redshift range $z=9$--$16$, with a uniform line-of-sight comoving distance interval of $3\,{\rm Mpc}$. Each lightcone spans an angular area of $6.5^\circ \times 6.5^\circ$ and
is sampled on a $400^2$ angular grid using the built-in lightcone
construction in \texttt{21cmFAST}, which follows the \texttt{cosmotile}\footnote{\url{https://cosmotile.readthedocs.io/en/stable/reference.html}}
algorithm to interpolate the simulation volume onto a past lightcone
geometry.

\begin{deluxetable}{lcc}
\tablecaption{Astrophysical parameters distinguishing the fiducial and early-heating scenarios. Only the three parameters listed here are varied between the two scenarios; all other astrophysical parameters are fixed.
\label{tab:Astrophysical_paras}}
\tablehead{
\colhead{Parameter} &
\colhead{Fiducial} &
\colhead{Early-heating}
}
\startdata
$\log_{10}(F_{\ast,10})$ & $-1.25$ & $-1.20$ \\
$\log_{10}(L_{\rm X}/{\rm SFR})$ & $40.50$ & $41.52$ \\
$E_0$ [eV] & $500$ & $300$
\enddata
\end{deluxetable}

In addition to these lightcones, we generate 30 smaller angular lightcones covering the broader redshift range $z=5.5$--$16$, again with a uniform line-of-sight comoving distance interval of $3\,{\rm Mpc}$. Each of these realizations spans $3.8^\circ \times 3.8^\circ$ and is sampled on a $285^2$ angular grid. These smaller lightcones are used exclusively to compute the angular auto-power spectrum of the patchy kSZ signal, $C_{\ell}^{\mathrm{kSZ}}|_{\rm EoR}$, which is required for constructing the kSZ filter described in subsection \ref{subsec:kSZ_Filter}. They are also used to illustrate the evolution of the volume-averaged neutral hydrogen fraction, the 21-cm differential brightness temperature, and  {the HI spin temperature}. We emphasize that these smaller lightcones are not used in the main $\mathrm{kSZ}^2 \times {\rm 21\,cm}^2$ cross-correlation analysis.  {Furthermore, the minimum multipole accessible from the smaller 
$3.8^\circ$ lightcones is $\ell_{\rm min} \sim 95$, fully covering  the multipole range $\ell \geq 100$ adopted in our analysis.}

The reionization and thermal histories of the simulations are governed by the underlying astrophysical parameters that control star formation and X-ray heating. In this work, we consider two scenarios: a fiducial model and an early-heating model characterized by earlier and more efficient X-ray heating. The fiducial model adopts the \textit{Munoz21} parameter template implemented in \texttt{21cmFASTv4} \citep{Munoz:2021psm}. The primary astrophysical parameters distinguishing the two models are summarized in Table \ref{tab:Astrophysical_paras}. Here, $F_{\ast,10}$ denotes the fraction of galactic gas converted into stars in halos with a mass of $10^{10}\,{\rm M}_\odot$, and is dimensionless. The parameter $L_{\rm X}/{\rm SFR}$ characterizes the specific X-ray luminosity produced per unit star formation rate that escapes the host galaxies, with units of $\mathrm{erg\,s^{-1}}/(\mathrm{M_\odot}\,\mathrm{yr}^{-1})$. Finally, $E_0$ represents the X-ray energy threshold above which photons can escape absorption by the host galaxies, and is given in units of $\mathrm{eV}$. We do not consider a late-heating scenario, since the kSZ signal is expected to remain weak over the redshift range of interest in such a model.

\begin{figure}[t]
    \centering
    \includegraphics[width=1.\linewidth]{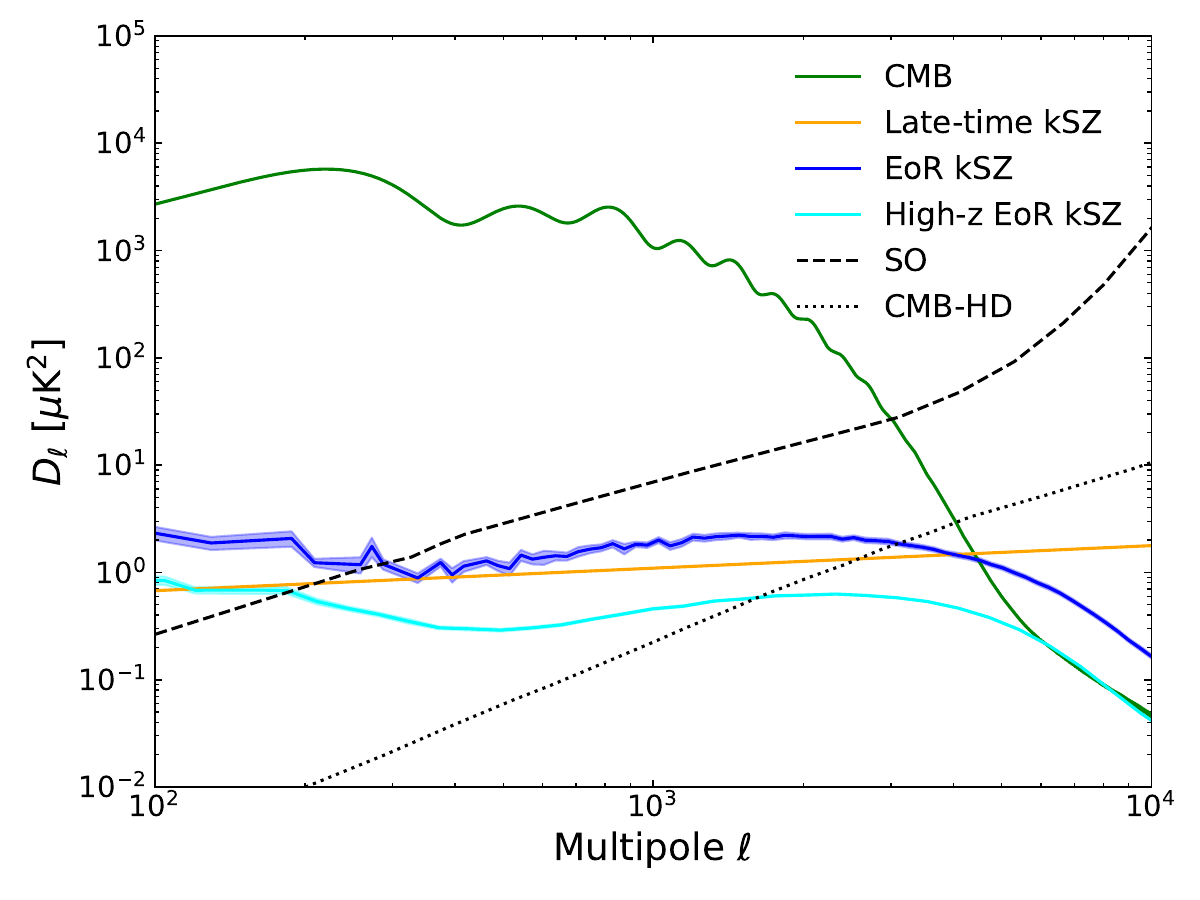}
    \caption{Angular power spectra of the components entering the kSZ filter defined in Eq.~(\ref{eq:kSZ_filter}). The high-$z$ EoR kSZ component integrated over $z=9$--$16$ is shown by the solid cyan curve, while the total EoR kSZ contribution integrated over $z=5.5$--$16$ is shown by the solid blue curve, as described in Sec.~\ref{subsec:lightcones}. The late-time kSZ power spectrum, shown by the solid orange curve, is modeled according to Eq.~(\ref{eq:latekSZ}), and the lensed primary CMB power spectrum, is shown by the solid green curve. The post-ILC residual noise power spectra, $N_\ell$, for the Simons Observatory and CMB-HD are shown by the dashed and dotted black curves, respectively. }
    \label{fig:kSZ_auto}
\end{figure}

\begin{figure*}[t]
    \centering
    \includegraphics[width=0.48\linewidth]{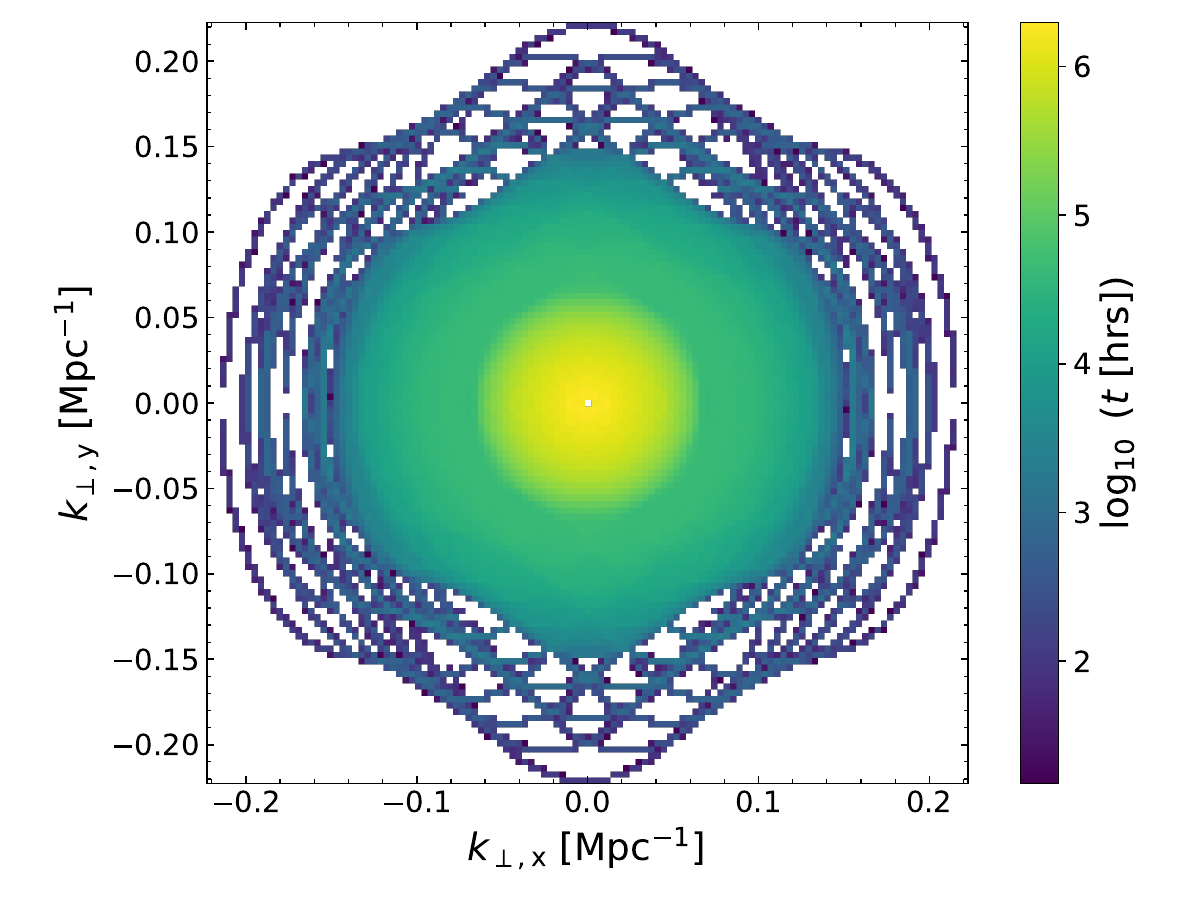}
    \includegraphics[width=0.48\linewidth]{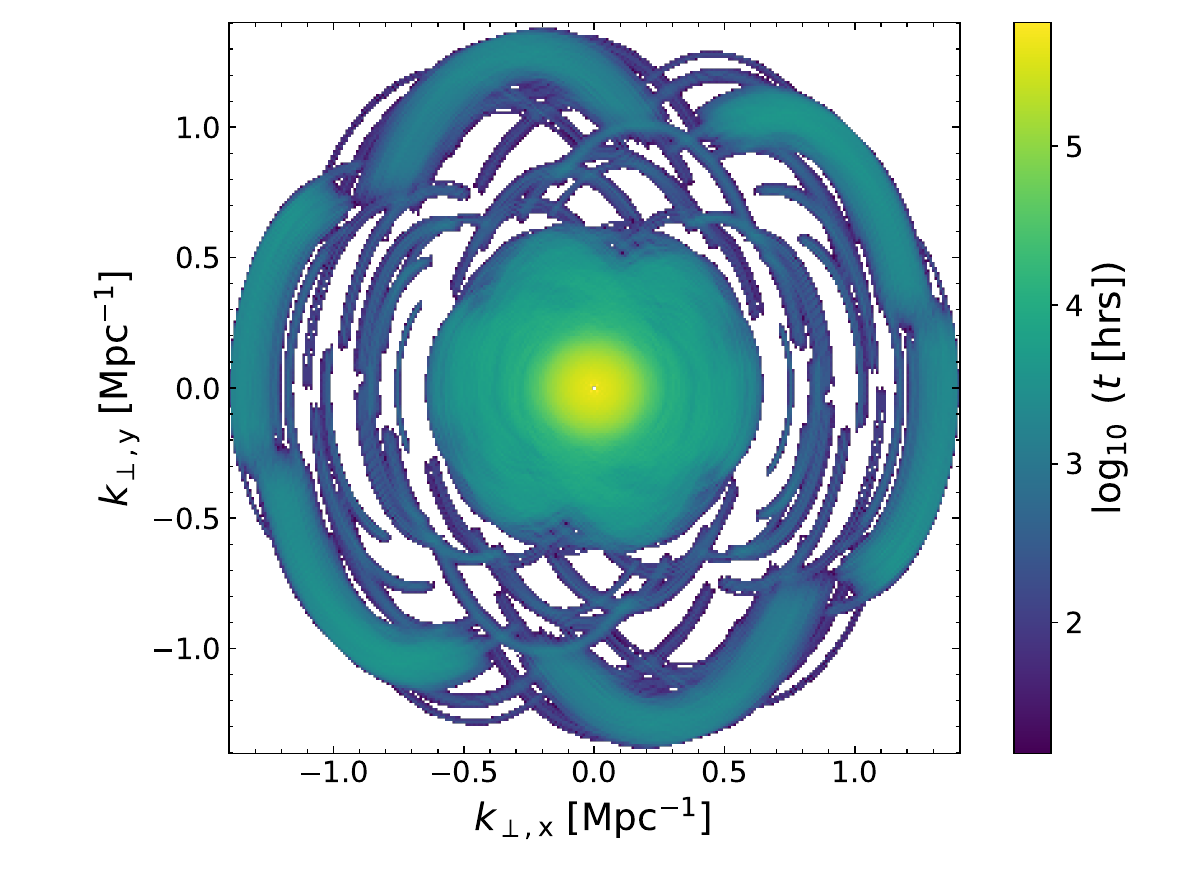}
    \caption{Effective integration time at $z=10$. \textit{Left}--For HERA configuration. \textit{Right}--For SKA-Low AA$^\ast$ configuration. The integration times are calculated with \texttt{21cmSense} and used to compute the thermal-noise power spectrum in Eq.~(\ref{eq:noise_ps}).
}
    \label{fig:IntegrationTime}
\end{figure*}

The evolution of the volume-averaged neutral fraction $\langle x_{\rm HI}\rangle$ 
\footnote{Here and throughout, $\langle\cdots\rangle$ denotes a volume average, 
and shaded regions in all figures indicate the $1\sigma$ uncertainty of the mean 
over realizations.}, the volume-averaged 21-cm differential brightness temperature 
$\langle \delta T_{\rm b}\rangle$, and the volume-averaged HI spin temperature $\langle T_{\rm s} \rangle$ for the two models are shown in the upper, middle, and lower panels of Fig.~\ref{fig:NeutralFraction_BrightnessTemperature}, respectively. The upper panel also includes observational constraints on the neutral hydrogen fraction. The lower panel additionally shows the CMB temperature $T_{\rm CMB}(z)$ as a dashed grey reference line. We compute the Thomson optical depth $\tau$ by integrating 
Eq.~(\ref{eq:tau}) from $z=0$ to the highest redshift of our simulations, $z=16$, assuming that the singly-ionized helium fraction traces that of ionized-hydrogen, $x_{\rm HeII}\simeq x_{\rm HII}$, as adopted in Eq.~(\ref{eq:ne}), and neglecting the doubly-ionized helium produced at lower redshifts. We obtain $\tau = 0.067$ and $0.072$ for the fiducial and early-heating models, respectively. Compared with the recent constraint $\tau = 0.058 \pm 0.006$ \citep{Tristram:2023haj}, both models are broadly consistent with the observational constraint, lying within $1.5\sigma$ and $2.3\sigma$ for the fiducial and early-heating models, respectively. We emphasize, however, that the two scenarios are not intended as best-fit models to all observational constraints, but rather to isolate the impact of different thermal histories on the cross-correlation signal. These mild deviations therefore do not affect the main conclusions of 
this work.

As an illustration of the simulated fields, Fig.~\ref{fig:kSZ_HI_maps} shows, in the left panel, one realization of the kSZ signal integrated over the redshift range $z=9$--$16$, which we hereafter refer to as the \textit{high-$z$ EoR kSZ} signal. The right panel shows a slice of the simulated 21-cm differential brightness temperature field at $z=10$. Both maps are generated using the fiducial model.

\subsection{Filters}
In this work, we apply a series of filters to the kSZ and 21-cm fields to account for foreground contamination, instrumental beam effect, and thermal noise. The filtering procedure for the two fields is designed to mimic the dominant limitations expected in realistic CMB and 21-cm observations.

\subsubsection{kSZ Filter}\label{subsec:kSZ_Filter}
For the kSZ field, we adopt a multipole-space filter following the approach developed in previous studies (e.g.,~\citealt{LaPlante:2022nlp}). The purpose of this filter is to preferentially retain the angular scales on which the high-$z$ EoR kSZ signal is expected to contribute significantly relative to the primary CMB, the late-time kSZ signal, residual foregrounds, and instrumental noise. We define the kSZ filter as
\begin{equation}\label{eq:kSZ_filter}
    F^{\mathrm{kSZ}}(\ell) = \frac{C_{\ell}^{\mathrm{kSZ}}|_{{\rm high}-z}}{C_{\ell}^{\mathrm{TT}} + C_{\ell}^{\mathrm{kSZ}}|_{\rm EoR} + C_{\ell}^{\mathrm{kSZ}}|_{\rm late} + N_{\ell}},
\end{equation}
where $C_{\ell}^{\mathrm{TT}}$ is the lensed primary CMB temperature power spectrum computed using \texttt{CAMB}\footnote{\url{https://camb.readthedocs.io/en/latest/}}, $C_{\ell}^{\mathrm{kSZ}}|_{\rm EoR}$ denotes the total kSZ power generated during the {\it entire} period of reionization ($5.5 \leq z \leq 16$), and $C_{\ell}^{\mathrm{kSZ}}|_{\rm late}$ represents the late-time kSZ contribution at $z\lesssim5.5$. The term $N_{\ell}$ accounts for the effective noise contribution from residual foregrounds and instrumental noise for a given CMB experiment.

The numerator, $C_{\ell}^{\mathrm{kSZ}}|_{{\rm high}-z}$, is the auto angular power spectrum of the high-$z$ EoR kSZ signal integrated over $z=9$--$16$, as defined in this work. In contrast, $C_{\ell}^{\mathrm{kSZ}}|_{\rm EoR}$ is computed from the kSZ signal integrated over $z=5.5$--$16$ using the smaller lightcone realizations described in Sec.~\ref{subsec:lightcones}. 

\begin{figure*}[t]
    \centering
    \includegraphics[width=0.48\linewidth]{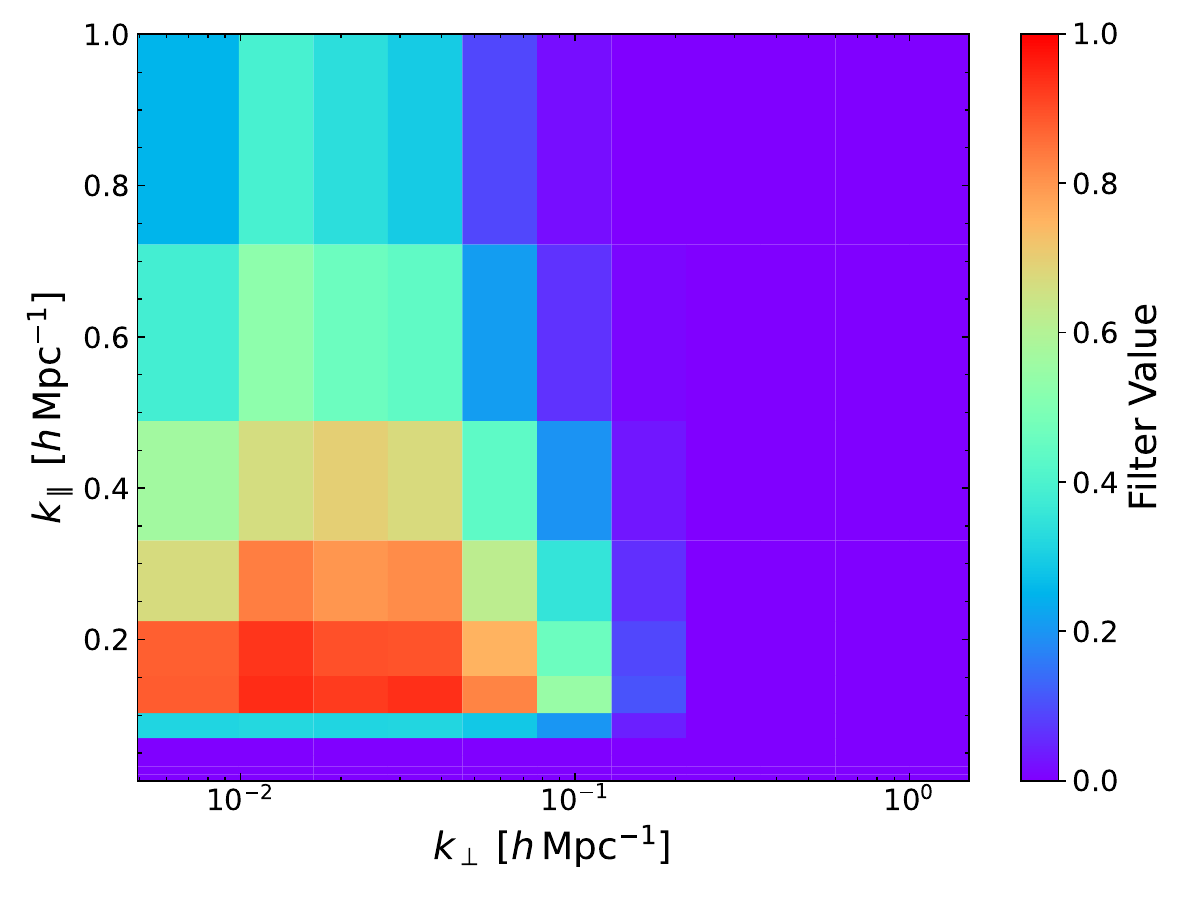}
    \includegraphics[width=0.48\linewidth]{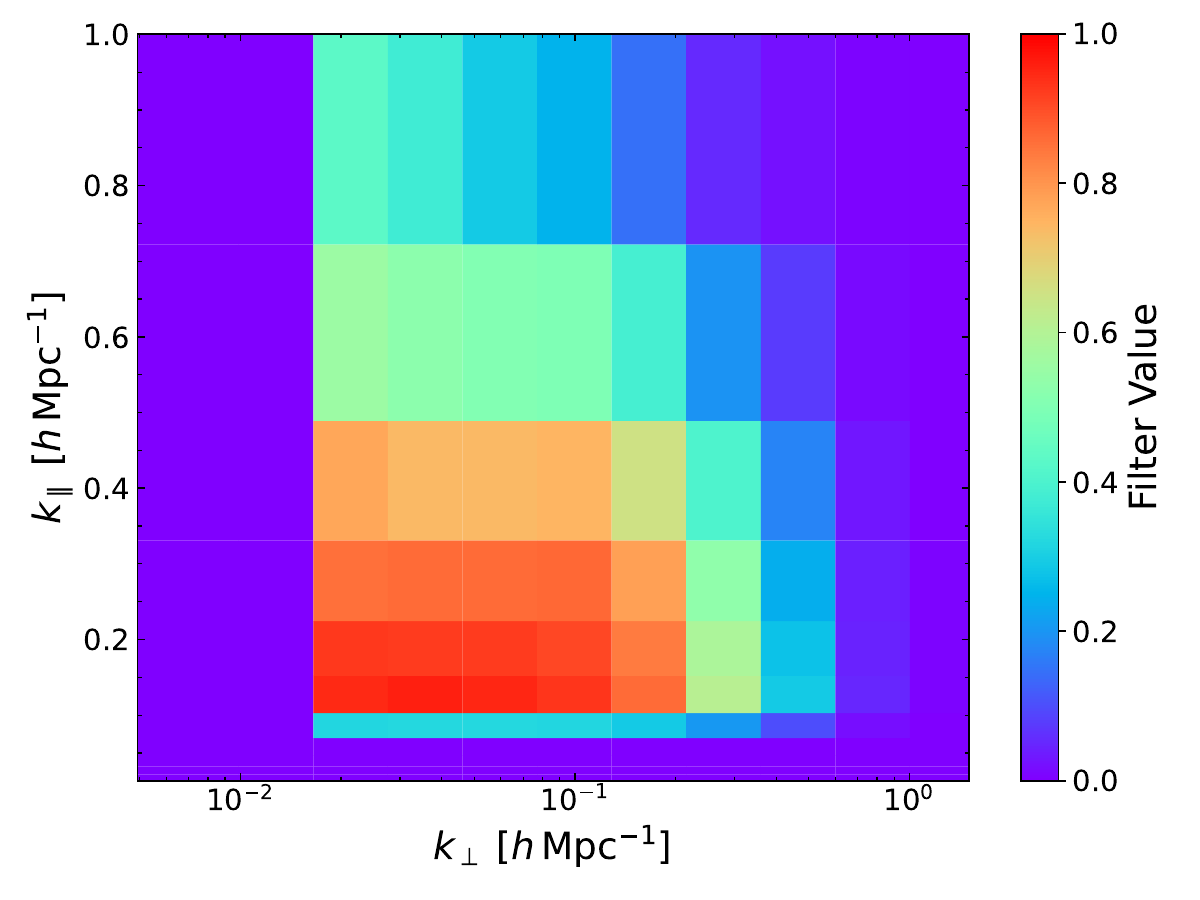}
    \caption{Combined Fourier-space filter applied to the simulated 21-cm fields centered at $z=10$ with chunk size of $\Delta z=2$. \textit{Left}--For HERA configuration. \textit{Right}--SKA-Low AA$^\ast$ configuration. The combined filter, defined in Eq.~(\ref{eq:HI_full_filter}), consists of an optimistic foreground-avoidance prescription, in which modes with $k_{\parallel}\leq 0.1\,h\,{\rm Mpc}^{-1}$ are excluded, and noise filtering.
}
    \label{fig:HI_Filter}
\end{figure*}

For the late-time kSZ contribution, we adopt the power-law fit from \citet{Park:2017amo}, which is given by
\begin{equation}\label{eq:latekSZ}
    D_{\ell}^{\rm kSZ,late} = 1.38\left(\frac{\ell}{3000}\right)^{0.21}\,{\rm \mu K}^{2},
\end{equation}
where $D_{\ell} = \ell (\ell + 1)C_{\ell}/(2\pi)$. The effective noise power spectrum $N_{\ell}$ is modeled using a realistic treatment of residual foreground and instrumental contamination. Specifically, we estimate the post-internal-linear-combination (ILC) noise obtained from a weighted combination of maps at different frequency channels. We use the publicly available \texttt{ORPHICS} package \footnote{\url{https://github.com/msyriac/orphics}} to perform the harmonic-space ILC and obtain the corresponding residual noise power spectra for the SO and CMB-HD experiments \citep{Hotinli:2021hih}. The individual components entering Eq.~(\ref{eq:kSZ_filter}) are shown in Fig.~\ref{fig:kSZ_auto}, where the high-$z$ and EoR kSZ power spectra are calculated from our fiducial model.  {Since the primary CMB dominates the denominator of Eq.~(\ref{eq:kSZ_filter}) by several orders of magnitude over all other contributions at low $\ell$, while residual foregrounds and instrumental noise dominate at high $\ell$ (see Fig.~\ref{fig:kSZ_auto}), the filter $F^{\rm kSZ}(\ell)$ strongly downweights both the low- and high-$\ell$ regimes and peaks at $3000 \lesssim \ell \lesssim 5000$. Consequently, uncertainties in $C_\ell^{\rm kSZ}|_{\rm EoR}$ in either of these regimes have a negligible impact on the construction and application of the filter.}

We further account for the finite angular resolution of the CMB experiments by applying a Gaussian beam window function
\begin{equation}
b(\ell)
=
\exp\left[
-\frac{\ell(\ell+1)\sigma_{\rm b}^{2}}{2}
\right],
\end{equation}
where
\begin{equation}
\sigma_{\rm b}
=
\frac{\theta_{\rm FWHM}}
{\sqrt{8\ln2}}.
\end{equation}
We take $\theta_{\rm FWHM} = 1.4'$ and $10''$ for the SO and CMB-HD experiments, respectively. The filtered kSZ field in harmonic space is then given by
\begin{equation}
 T^{\rm (f)}_{\mathrm{kSZ}}(\ell)
=
F_{\mathrm{kSZ}}(\ell)
b(\ell)
\Delta T_{\mathrm{kSZ}}(\ell).
\end{equation}

\subsubsection{21-cm Foreground and Noise Filters}

Foreground contamination poses a particularly severe challenge for 21-cm observations, with foreground emission exceeding the cosmological signal by several orders of magnitude. Rather than attempting to model and subtract these foregrounds explicitly, we adopt a foreground-avoidance approach, in which Fourier modes expected to be strongly contaminated are excluded from the analysis.

We consider two foreground-avoidance scenarios. In an optimistic scenario, we assume that the chromatic response of the interferometer can be well calibrated and that foreground contamination is confined to the lowest $k_\parallel$ modes. We therefore apply a simple high-pass filter
\begin{equation}
f_{\mathrm{fore,opti}}^{\mathrm{21cm}}(k_{\perp}, k_{\parallel}) =
\begin{cases}
1 & k_{\parallel} > k_{\parallel,0}, \\
0 & \mathrm{otherwise},
\end{cases}
\end{equation}
where $k_{\parallel,0}$ denotes the minimum line-of-sight wavenumber retained in the analysis.

A more realistic treatment accounts for the chromatic response of an
interferometer, which couples angular and frequency-dependent structure and
causes spectrally smooth foregrounds to leak into higher-$k_\parallel$ modes.
This mode mixing produces the characteristic foreground wedge in cylindrical
Fourier space. In the idealised case in which the instrument itself introduces
no additional spectral structure and contributions from RFI and polarisation
leakage are neglected, the contaminated region is bounded by the so-called
\emph{horizon wedge}
\begin{equation}
k_{\parallel} < m(z) k_{\perp},
\end{equation}
with the horizon wedge slope
\begin{equation}
m(z)=\frac{H(z)\chi(z)}{c(1+z)}.
\end{equation}
Here, $\chi(z)$ is the comoving radial distance. The corresponding foreground-avoidance filter is then
\begin{equation}
f_{\mathrm{fore,real}}^{\mathrm{21cm}}(k_{\perp}, k_{\parallel}) =
\begin{cases}
1 & k_{\parallel} > m(z) k_{\perp}, \\
0 & \mathrm{otherwise}.
\end{cases}
\end{equation}

In addition to foreground avoidance, we account for the loss of sensitivity caused by instrumental thermal noise using a Wiener filter
\begin{equation}
f_{\mathrm{noise}}^{\mathrm{21cm}}(k_{\perp}, k_{\parallel}) =
\frac{P^{\mathrm{21cm}}(k_{\perp}, k_{\parallel})}{P^{\mathrm{21cm}}(k_{\perp}, k_{\parallel}) + P^{\mathrm{21cm}}_{\rm noise}(k_{\perp})},
\end{equation}
where $P^{\mathrm{21cm}}$ is the cosmological 21-cm auto-power spectrum and $P^{\mathrm{21cm}}_{\rm noise}$ is the corresponding thermal-noise power spectrum. We adopt \citep{Liu:2019awk} 
\begin{equation}\label{eq:noise_ps}
    P^{\rm 21cm}_{\rm noise} = \frac{T_{\rm sys}^{2}\Omega_{\rm p}^{2}X^{2}(\nu)Y(\nu)}{\Omega_{\rm pp}N_{\rm pol}t_{u}},
\end{equation}
where
\begin{equation}
T_{\rm sys}\,[{\rm K}] = 260\,{\rm K}\left(\frac{\nu}{150\,{\rm MHz}}\right)^{-2.60} + T_{\rm rcv},
\end{equation}
is the total system temperature of the interferometer at frequency $\nu$ \citep{Eastwood:2019rwh}. Here, $\Omega_{\rm p}$ is the integral of the primary beam, $\Omega_{\rm pp}$ is the integral of the squared primary beam, and $N_{\rm pol}=2$ is the number of independent polarization states. The factors $X(\nu)$ and $Y(\nu)$ convert angular and frequency intervals into comoving distances and are defined as
\begin{equation}
\begin{aligned}
X(\nu)
&\equiv
\frac{{\rm d}r_{\perp}}{{\rm d}\theta}
=
D_{\perp}(\nu),\\
Y(\nu)
&\equiv
\frac{{\rm d}r_{\parallel}}{{\rm d}\nu}
=
\frac{c(1+z)}
{H(z)\nu},
\end{aligned}
\end{equation}
where $D_{\perp}$ is the transverse comoving distance. For a spatially flat Universe, $D_{\perp}=\chi$.

The quantity $t_u$ represents the effective integration time accumulated in a given $uv$ cell and therefore depends on the baseline distribution, observing strategy, and Earth-rotation synthesis of the array. We calculate the corresponding $uv$-dependent integration time distributions for both HERA and SKA-Low AA$^{*}$ using the publicly available \texttt{21cmSense} package. We assume a tracked observing strategy with $t_{\rm tracking}=6\,{\rm h\,day^{-1}}$, an integration time of $t_{\rm int}=60\,{\rm s}$ per visibility, and a total observing duration of five years, corresponding to $180\times5$ observing days.

For computational convenience, we restrict the maximum baseline of SKA-Low AA$^{*}$ included in the sensitivity calculation to $b_{\rm max}=5\,{\rm km}$. This approximation is sufficient for the angular scales relevant to our analysis, while the antennas at larger separations provide comparatively sparse sampling. The resulting integration time distributions for HERA (left) and SKA-Low AA$^{*}$ (right) as a function of $k_\perp$ at $z=10$ are shown in Fig.~\ref{fig:IntegrationTime}.

\begin{figure}
    \centering
    \includegraphics[width=1.\linewidth]{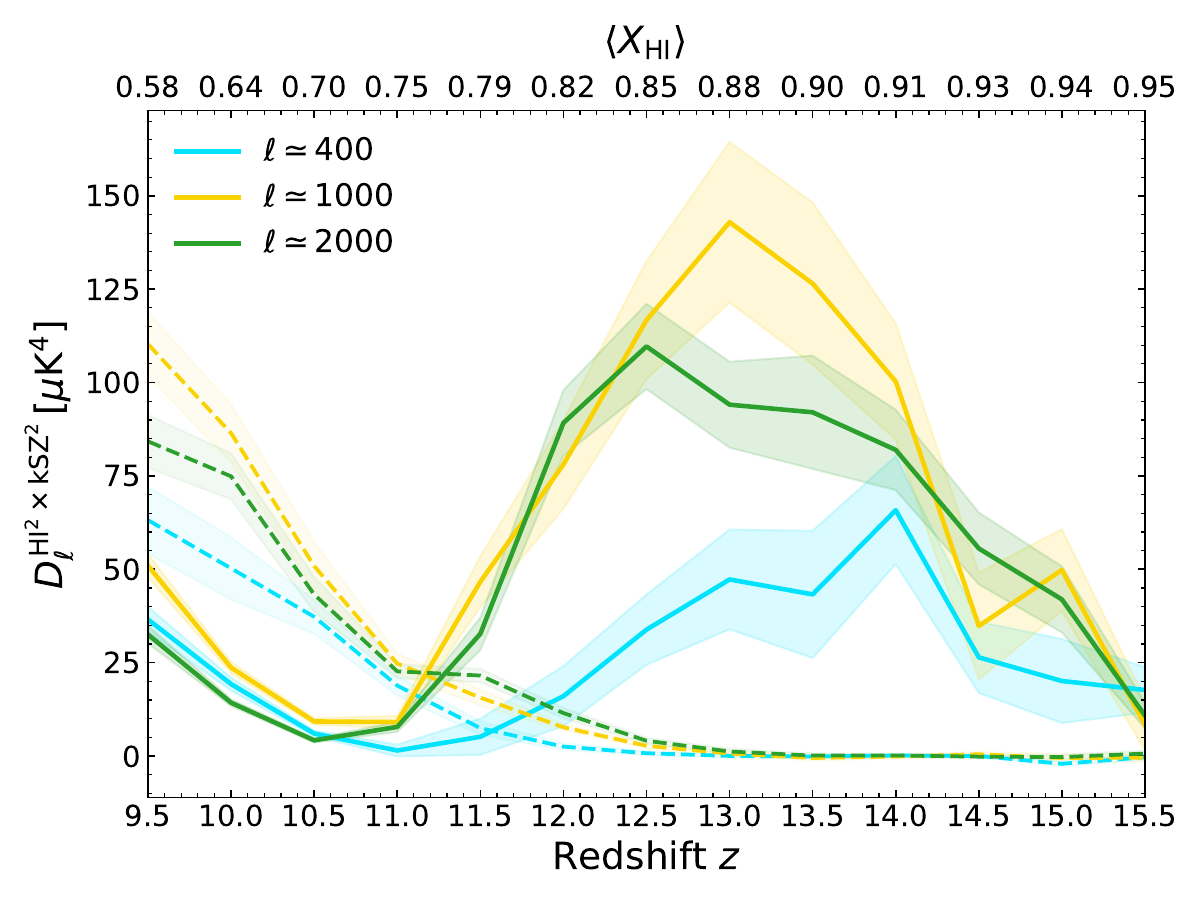}
    \caption{Redshift evolution of the $\mathrm{kSZ}^2\times21\mathrm{cm}^2$ power spectrum ($D_{\ell}\equiv \ell(\ell +1)C_{\ell}/2\pi$) at representative multipoles in the fiducial model. Solid curves include spatial fluctuations in the HI spin temperature, while dashed curves adopt the saturated-heating approximation $T_{\rm s}\gg T_{\rm CMB}$. The 21-cm field is evaluated in redshift chunks of width $\Delta z=1$, with foreground avoidance excluding modes with $ k_{\parallel,0}=0.1\,h\,{\rm Mpc}^{-1}$ and instrumental-noise filtering for the SKA-Low AA$^\ast$ configuration. The kSZ field is filtered using the CMB-HD configuration.}
    \label{fig:Redshift_Evolution_Ts_vs_noTs}
\end{figure}

The combined 21-cm filter is therefore
\begin{equation}\label{eq:HI_full_filter}
f_{\mathrm{21cm}}(\mathbf{k}) = f_{\mathrm{fore}}^{\mathrm{21cm}}(\mathbf{k}) \, f_{\mathrm{noise}}^{\mathrm{21cm}}(\mathbf{k}),
\end{equation}
where $f_{\mathrm{fore}}^{\mathrm{21cm}}$ is chosen according to either the optimistic or realistic foreground-avoidance filter described above. The filtered 21-cm field is then
\begin{equation}
    T^{\rm (f)}_{21\mathrm{cm}}(\mathbf{k}) = f_{\mathrm{21cm}}(\mathbf{k}) \delta T_{\rm b}(\mathbf{k}).
\end{equation}

\begin{figure*}[t]
    \centering
    \includegraphics[width=0.325\linewidth]{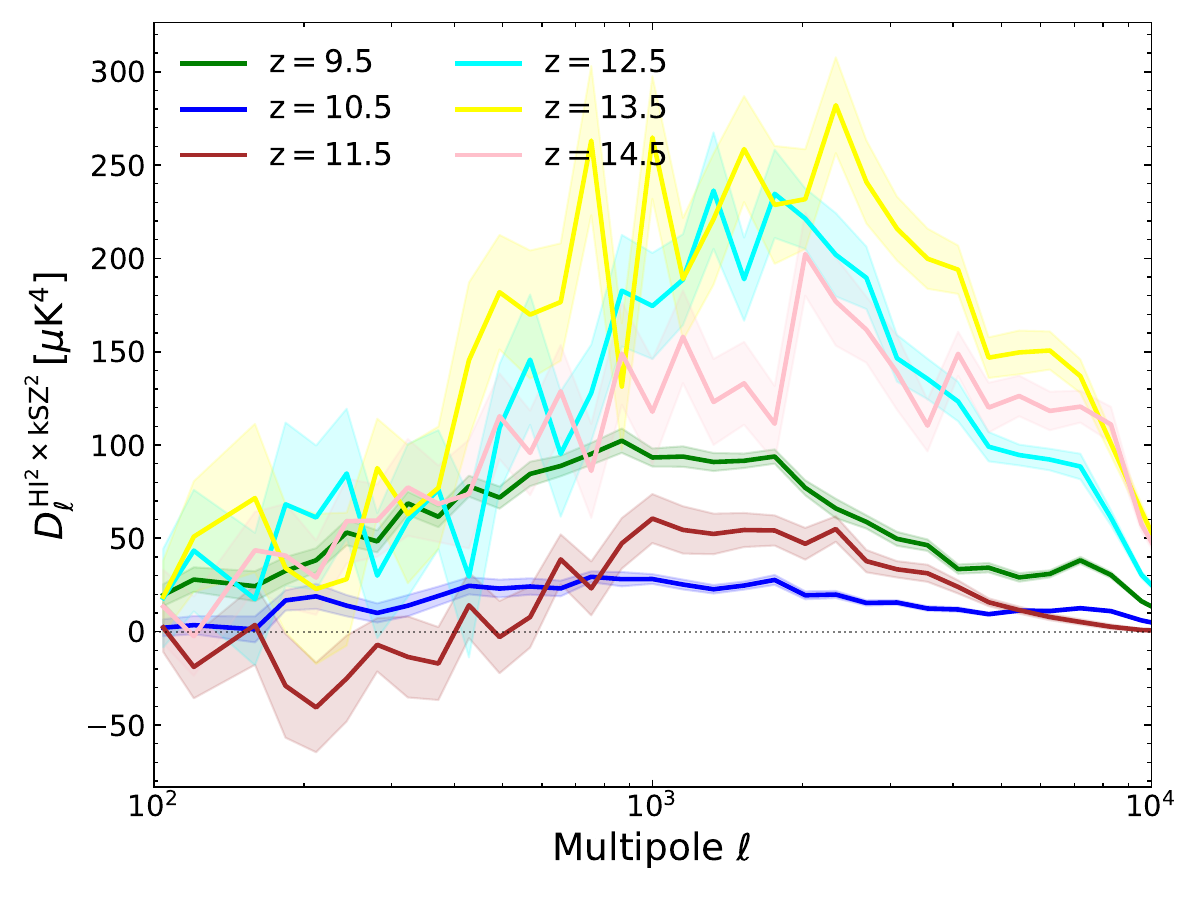}
    \includegraphics[width=0.325\linewidth]{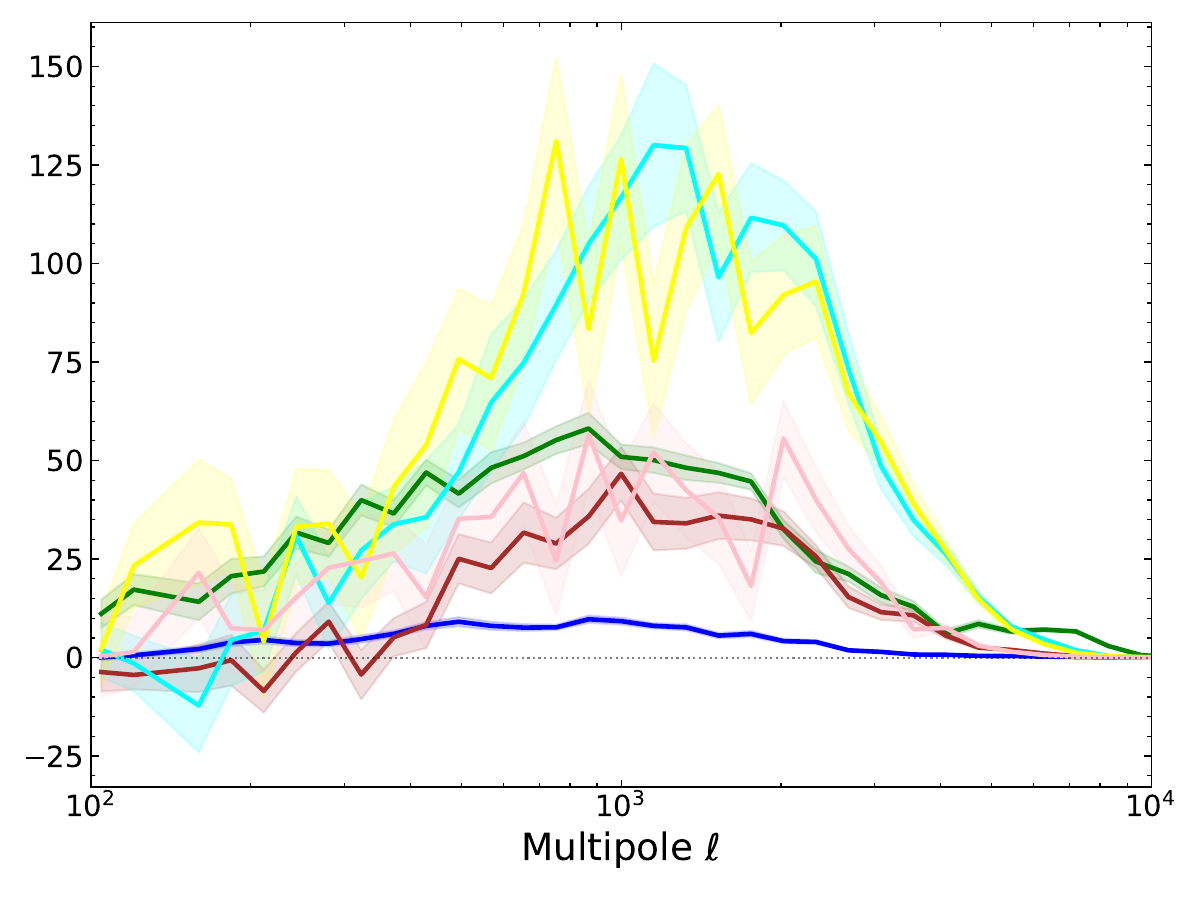}
    \includegraphics[width=0.325\linewidth]{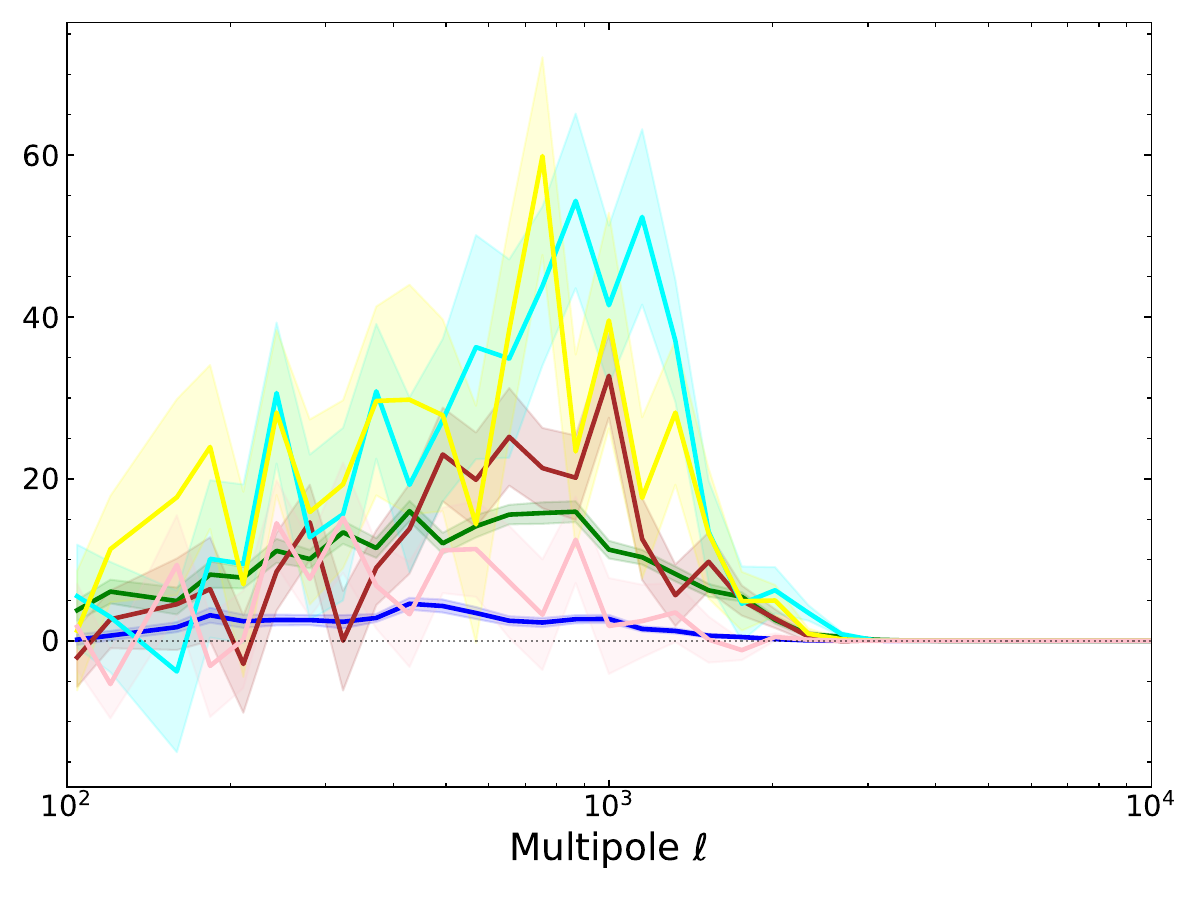}
    
    \caption{Angular $\mathrm{kSZ}^2\times21\mathrm{cm}^2$ cross-power spectra for redshift chunks of width $\Delta z=1$ in the fiducial model. \textit{Left}--The 21-cm field is filtered using foreground avoidance with $k_{\parallel,0}=0.1\,h\,{\rm Mpc}^{-1}$ alone. \textit{Middle:} The same foreground filter is combined with the instrumental-noise filter for SKA-Low AA$^\ast$. \textit{Right}--The same foreground filter is combined with the instrumental-noise filter for HERA. The kSZ field is filtered using the CMB-HD configuration.}
    \label{fig:APS_different_HI_Filter}
\end{figure*}

For the noise power spectrum calculation, we adopt the receiver temperature $T_{\rm rcv}=100\,{\rm K}$ and $40\,{\rm K}$ for HERA \citep{Zhao:2022ren} and SKA-Low AA$^{*}$ \citep{SKA:2018ckk}, respectively. The resulting two-dimensional 21-cm filters of HERA (left) and SKA-Low AA$^{*}$  (right) in the fiducial model are shown in Fig.~\ref{fig:HI_Filter}. For illustration, we consider a redshift chunk centered at $z=10$ with a width of $\Delta z=2$ and adopt the optimistic foreground-avoidance model with $k_{\parallel,0}=0.1\,h\,{\rm Mpc}^{-1}$. The different configurations of the two arrays lead to different regions of Fourier-space sensitivity. HERA, owing to its highly compact configuration with a minimum baseline of approximately $14\,{\rm m}$, provides strong sensitivity to relatively large transverse scales. SKA-Low AA$^{*}$, by contrast, has a more extended configuration with a minimum baseline of approximately $42\,{\rm m}$, giving it enhanced sensitivity at intermediate and smaller transverse scales relative to HERA. This complementarity is important for determining the range of spatial modes contributing to the $\mathrm{kSZ}^2\times\mathrm{21\,cm}^2$ cross-correlation.

\section{Cross-Correlation Signals}\label{sec:CAPS}
We extract a redshift chunk of width $\Delta z$ from the 21-cm differential brightness temperature lightcone and apply the corresponding Fourier-space filters. The filtered field is then transformed back to configuration space and squared. The squared 21-cm field is subsequently projected along the line of sight to construct a two-dimensional map. In parallel, we apply the kSZ filter to the high-$z$ EoR kSZ map and square it. Finally, we compute the angular cross-power spectrum $C_{\ell}^{\mathrm{kSZ}^2\times21\mathrm{cm}^2}$ between these maps.

\subsection{Impact of the HI Spin Temperature}\label{subsec:SpinTemperature}
A key aspect of this work is the inclusion of spatial fluctuations in the HI spin temperature during the early stages of the EoR. To quantify their impact on the $\mathrm{kSZ}^2\times21\mathrm{cm}^2$ cross-correlation, we compute the cross-correlation using the full 21-cm signal, which retains the HI spin-temperature fluctuations, and compare it with the commonly adopted saturated-heating approximation in which $T_{\rm s}\gg T_{\rm CMB}$ and the factor
\begin{equation}
    1-\frac{T_{\rm CMB}(z)}{T_{\rm s}(z,\hat r)}
\end{equation}
in Eq.~(\ref{eq:HI}) is set to unity.

Figure~\ref{fig:Redshift_Evolution_Ts_vs_noTs} shows the resulting angular cross-correlation signal, $C_{\ell}^{\mathrm{kSZ}^2\times21\mathrm{cm}^2}(z)$, at several representative multipoles as a function of redshift for the fiducial reionization scenario. We consider redshift chunks of width $\Delta z=1$ (spanning a representative frequency bandwidth of $\sim 12\,{\rm MHz}$ at $z=10$) and adopt CMB-HD and SKA-Low AA$^\ast$ for the kSZ and 21-cm observations, respectively. For better visualization, we consider an optimistic foreground-avoidance scenario, in which modes with $k_{\parallel} \leq 0.1\,h\,{\rm Mpc}^{-1}$ are excluded. The solid curves include the spatially varying spin-temperature field, while the dashed curves adopt the saturated-heating approximation.

The difference between the two cases becomes particularly pronounced during the early stages of the EoR, when the thermal state of the IGM is still evolving. Under the saturated-heating approximation, the cross-correlation signal decreases rapidly towards higher redshifts and becomes nearly negligible at $z\gtrsim12$; in this limit, the 21-cm signal is determined primarily by the neutral hydrogen abundance and the matter density.  {This behaviour is broadly consistent with the results of \cite{Zhou:2025fgv}, who also find a decreasing cross-correlation signal with increasing redshift over $z \sim 9$--$11$ under the saturated-heating approximation, albeit with different simulations and experimental configurations.}

When spin-temperature fluctuations are included, the redshift evolution of the cross-correlation changes substantially. As the redshift increases, the signal first decreases toward a pronounced minimum, then rises, and eventually declines again at the highest redshifts. The initial suppression occurs as the spin temperature approaches the CMB temperature, causing the factor $1-T_{\rm CMB}/T_{\rm s}$ to approach zero. The 21-cm differential brightness temperature is therefore strongly suppressed, and consequently so is the $\mathrm{kSZ}^2\times21\mathrm{cm}^2$ cross-correlation. At higher redshifts, the IGM enters the absorption regime with $T_{\rm s}<T_{\rm CMB}$, while spatial fluctuations in $T_{\rm s}$ become increasingly important in shaping the morphology and amplitude of the 21-cm field. These thermal fluctuations enhance the cross-correlation, causing the signal to rise. At sufficiently high redshifts ($z\gtrsim15.5$), however, the IGM is predominantly neutral and ionized regions are rare, causing the kSZ contribution to diminish rapidly and the cross-correlation to become weak again.

\begin{figure*}[t]
    \centering
    \includegraphics[width=0.48 \linewidth]{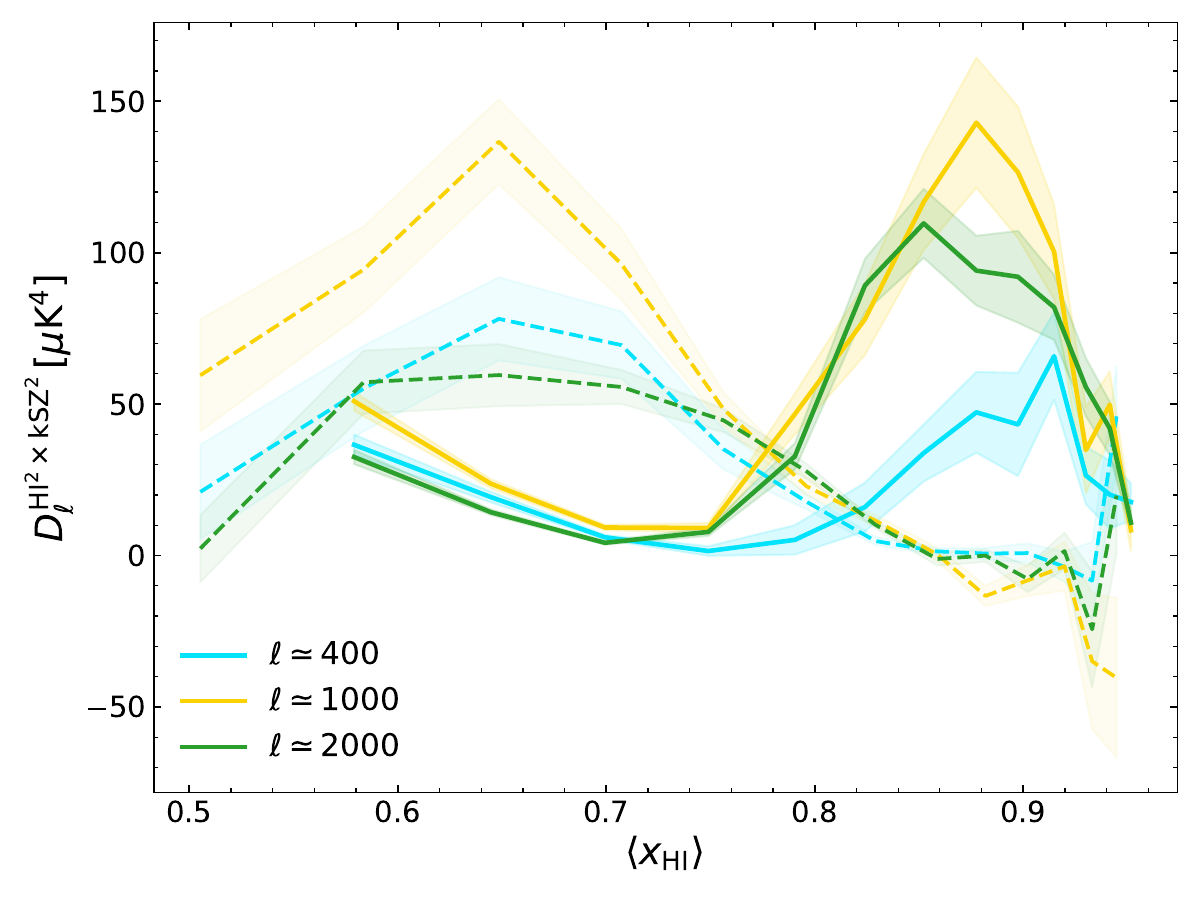}
    \includegraphics[width=0.48 \linewidth]{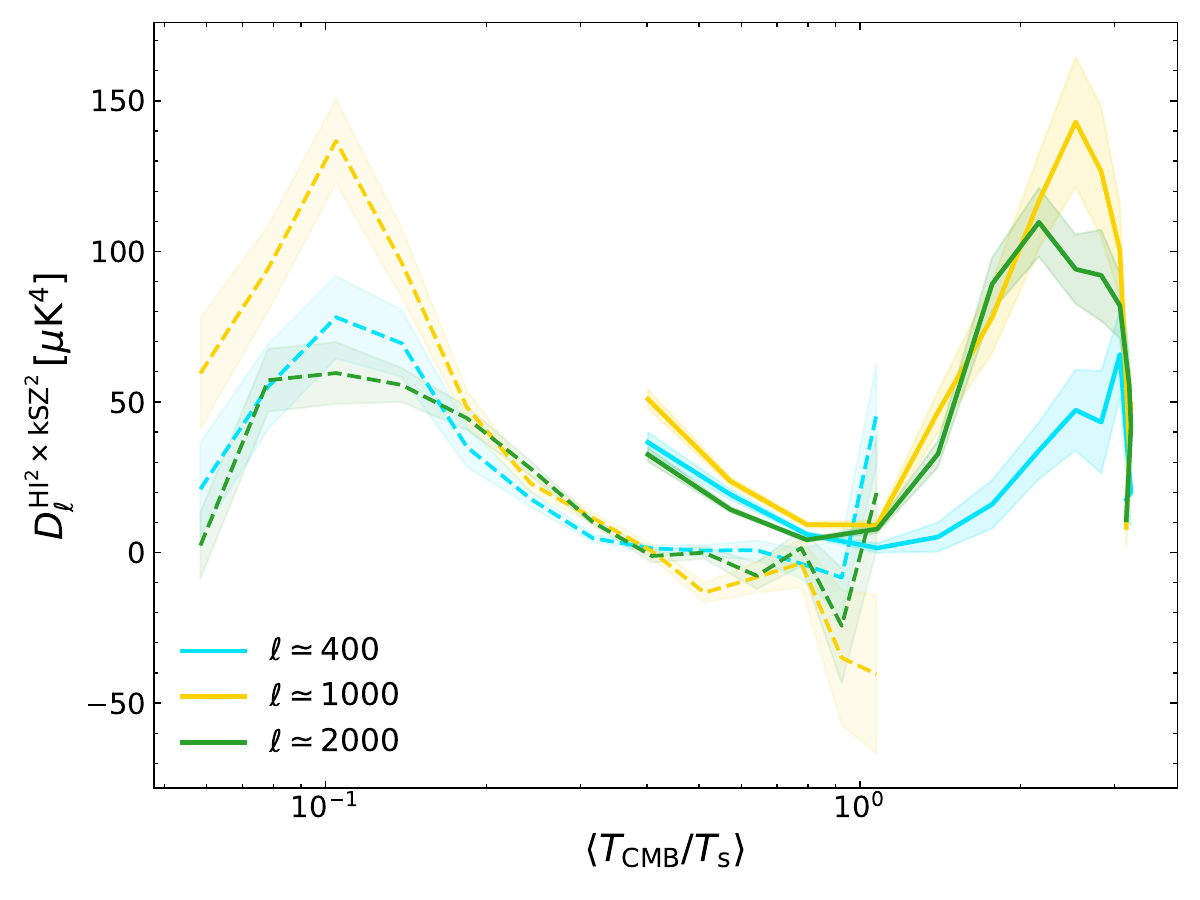}
    \caption{The $\mathrm{kSZ}^2\times21\mathrm{cm}^2$ cross-correlation as a function of the volume-averaged neutral fraction $\langle x_{\rm HI}\rangle$ and the ratio $\langle T_{\rm CMB}/T_{\rm s}\rangle$ for the fiducial (solid) and early-heating (dashed) scenarios. \textit{Left}--Cross-correlation as a function of $\langle x_{\rm HI}\rangle$. \textit{Right}--Cross-correlation as a function of $\langle T_{\rm CMB}/T_{\rm s}\rangle$. The cross-correlation is evaluated in redshift chunks of width $\Delta z=1$, assuming CMB-HD and SKA-Low AA$^\ast$ for the kSZ and 21-cm observations, respectively. An optimistic foreground-avoidance filter with $k_{\parallel,0}=0.1\,h\,{\rm Mpc}^{-1}$ is applied to the 21-cm field.
}
\label{fig:DifferentReionization_history}
\end{figure*}

Interestingly, at redshifts $z\lesssim 11$, the cross-correlation signal predicted under the saturated-heating approximation has a systematically higher amplitude than that obtained when spin-temperature fluctuations are included. The quantitative importance of this difference can be understood from the redshift evolution of the volume-averaged $\langle T_{\rm CMB}/T_{\rm s}\rangle$. In the fiducial model, we find
\begin{equation}
    \left \langle \frac{T_{\rm CMB}}{T_{\rm s}}\right\rangle_{z=9} \simeq 0.27
\end{equation}
at $z=9$. Thus, even at this relatively late stage, the assumption $T_{\rm s}\gg T_{\rm CMB}$ is not strictly valid. Because the observable considered here involves the squared 21-cm field, the effect of the multiplicative factor $1-T_{\rm CMB}/T_{\rm s}$ is amplified further. Under the saturated-heating approximation, this factor is set to unity, thereby increasing the amplitude of the predicted $\mathrm{kSZ}^2\times21\mathrm{cm}^2$ cross-correlation. The discrepancy between the two predictions therefore demonstrates that the thermal state of the IGM cannot, in general, be neglected when interpreting the cross-correlation during the earlier stages of the EoR.

We caution, however, that $\langle T_{\rm CMB}/T_{\rm s}\rangle$ should be regarded only as a rough indicator of the thermal state of the IGM. The cross-correlation statistic considered here depends on fluctuations in the 21-cm signal and their correlation with the kSZ effect, rather than on the volume-averaged ratio $\langle T_{\rm CMB}/T_{\rm s}\rangle$ itself. It is therefore best interpreted as a loose diagnostic of the physical conditions, rather than as a quantity that directly determines the amplitude of the cross-correlation.

\subsection{Dependence on the 21-cm Filter}\label{subsec:HI_Filter}
The $\mathrm{kSZ}^2\times21\mathrm{cm}^2$ cross-correlation is also strongly sensitive to the Fourier-space filtering of the 21-cm field. This dependence arises from two complementary effects. First, foreground avoidance removes a region of $(k_\perp,k_\parallel)$ space dominated by spectrally smooth foreground emission. Second, the instrumental noise of an interferometer imposes a scale-dependent Wiener suppression whose form is determined by the array configuration and its $uv$ sampling. Since the filtering is applied before squaring the 21-cm field, these effects modify not only the overall amplitude but also the distribution of Fourier-mode pairs contributing to the projected $21\mathrm{cm}^2$ field. Consequently, the shape of the resulting angular cross-power spectrum can be substantially altered.

\begin{figure*}[t]
    \centering
    \includegraphics[width=0.48 \linewidth]{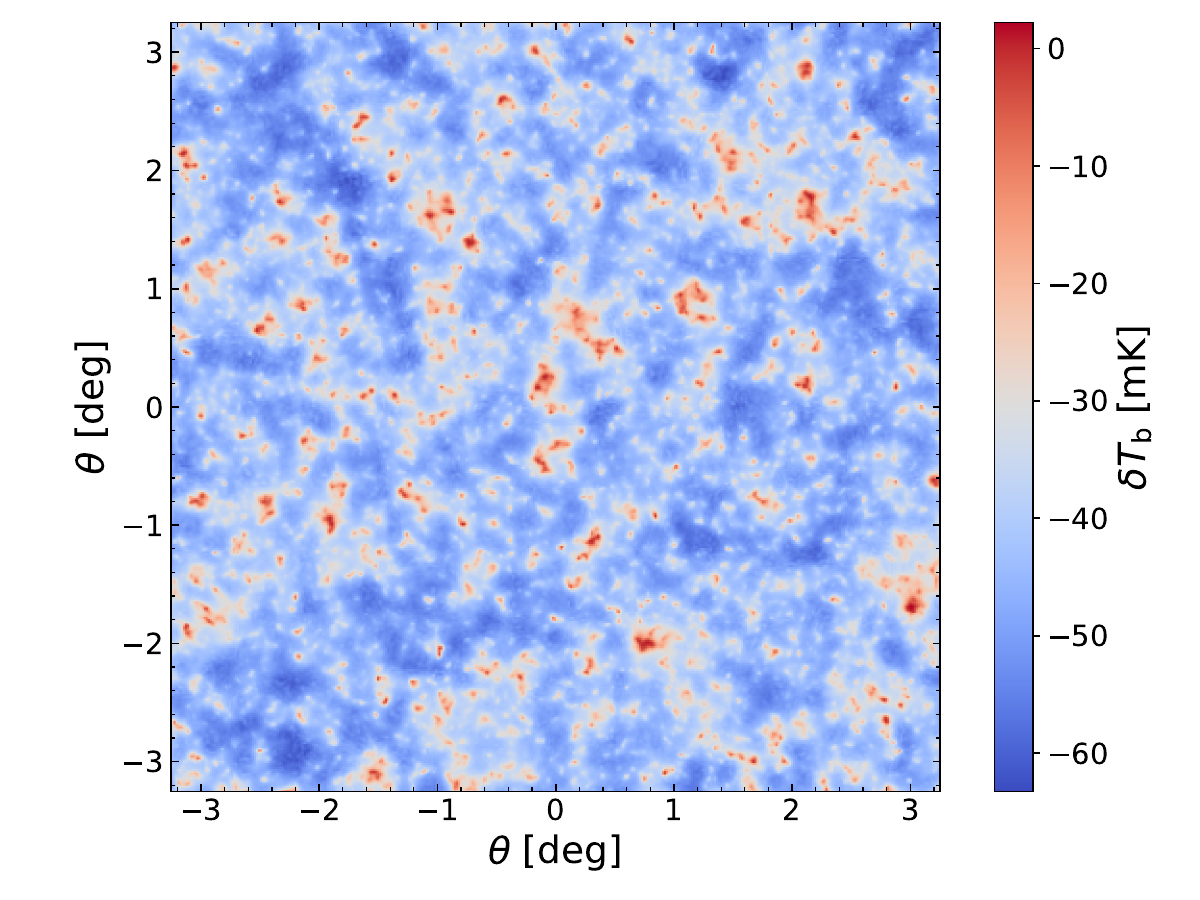}
    \includegraphics[width=0.48 \linewidth]{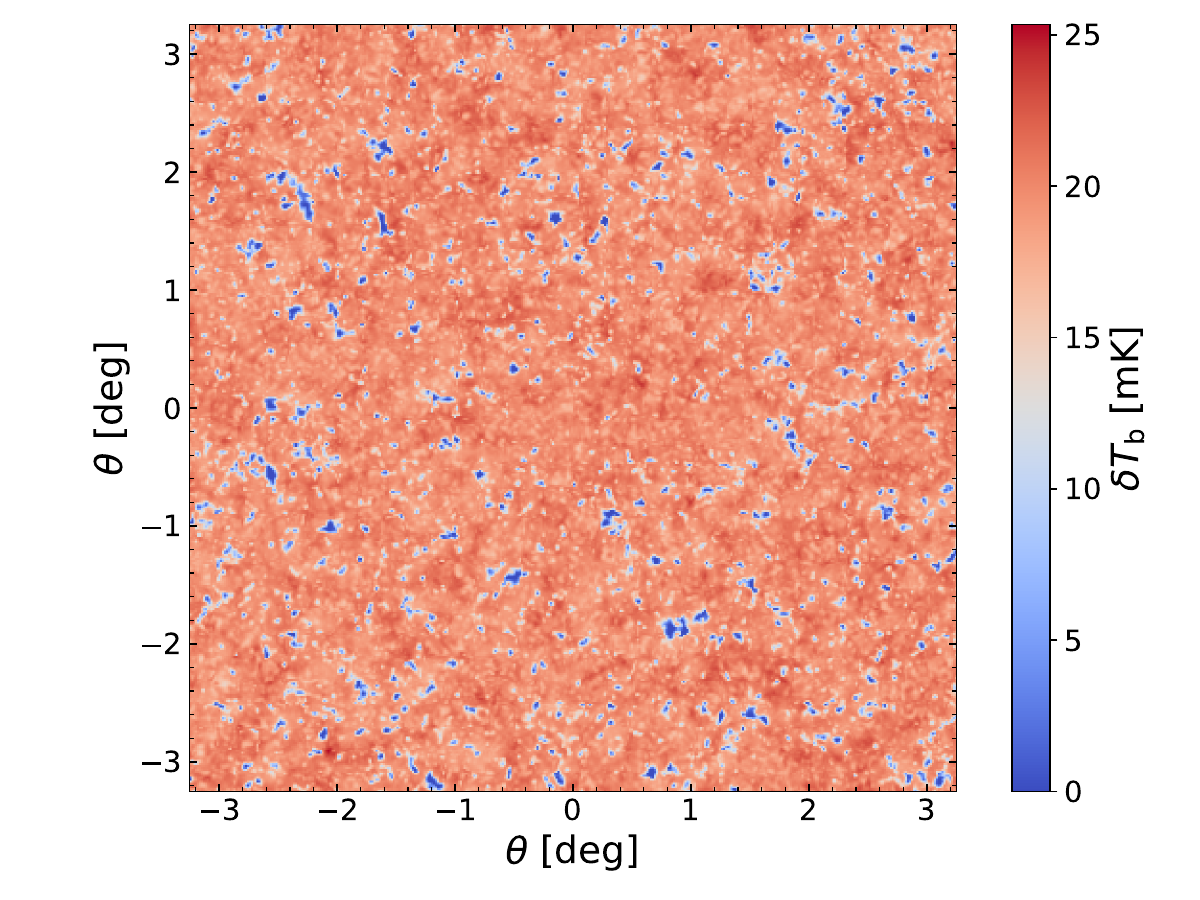}
    \caption{Slices of the 21-cm differential brightness temperature at the epoch
    where the $\mathrm{kSZ}^{2}\times21\,\mathrm{cm}^{2}$ cross-correlation peaks, illustrating the physical origin of the peaks in the right panel of Fig.~\ref{fig:DifferentReionization_history}.
    \textit{Left}--fiducial model, for which the cross-correlation peaks at $\langle T_{\rm CMB}/T_{\rm s}\rangle \simeq 2.5$. \textit{Right}--early-heating model, for which the cross-correlation peaks at $\langle T_{\rm CMB}/T_{\rm s}\rangle \simeq 0.1$.
}
\label{fig:HI_cmbz_Ts}
\end{figure*}

To illustrate this effect, Fig.~\ref{fig:APS_different_HI_Filter} shows $C_{\ell}^{\mathrm{kSZ}^2\times21\mathrm{cm}^2}$ as a function of multipole for several redshift bins in the fiducial model. Each redshift chunk has width $\Delta z=1$. The kSZ field is filtered using the CMB-HD configuration, while three different treatments of the 21-cm field are considered. The left panel includes only the optimistic foreground-avoidance filter, with modes satisfying $k_\parallel\leq 0.1\,h\,{\rm Mpc}^{-1}$ removed. The middle panel additionally applies the SKA-Low AA$^\ast$ instrumental-noise filter, while the right panel applies the corresponding HERA noise filter.

The comparison demonstrates that the impact of the 21-cm instrument is strongly redshift dependent. At relatively low redshifts, $z\lesssim11$, the cross-correlation receives a considerable contribution from intermediate scales. By this stage of the EoR, ionized regions have grown sufficiently large that the characteristic spatial structure of the ionization field is shifted towards larger physical scales. Both HERA and SKA-Low AA$^\ast$ retain appreciable sensitivity to these modes, preserving the spectral shape, albeit with a suppressed amplitude.

At higher redshifts, $z\gtrsim11$, the situation changes significantly. The ionized regions are smaller, and the spatial fluctuations associated with the first ionizing sources increasingly contribute on smaller scales. The more extended baseline distribution of SKA-Low AA$^\ast$ provides markedly better sensitivity to these smaller scales than the more compact HERA configuration. As a result, the SKA-Low AA$^\ast$-filtered maps retain a recognizable cross-correlation spectrum at these higher redshifts, whereas the HERA-filtered signals become strongly suppressed and are associated with substantially larger uncertainties.

This difference illustrates that the detectability of the $\mathrm{kSZ}^2\times21\mathrm{cm}^2$ signal is determined not only by the total sensitivity of a 21-cm experiment, but also by the range of Fourier modes to which the array is sensitive. In particular, the evolving characteristic scale of the ionization structure to be detected during the EoR must be matched by the Fourier coverage of the experiment.

A notable feature is the strong suppression of the SKA-Low AA$^\ast$ signal around $z=14.5$. In our calculation, the maximum baseline of SKA-Low AA$^\ast$ is truncated to $5\,{\rm km}$ for computational convenience. This truncation limits the maximum accessible transverse wavenumber and therefore removes part of the small-scale information that becomes increasingly important at these high redshifts. We verify this interpretation by extending the maximum baseline in an additional calculation, which restores part of the suppressed cross-correlation signal. This result indicates that the apparent loss of sensitivity at the highest redshifts is, at least in part, a consequence of the imposed baseline truncation rather than an intrinsic limitation of the SKA-Low AA$^\ast$ concept.

\begin{figure}
    \centering
    \includegraphics[width=1.\linewidth]{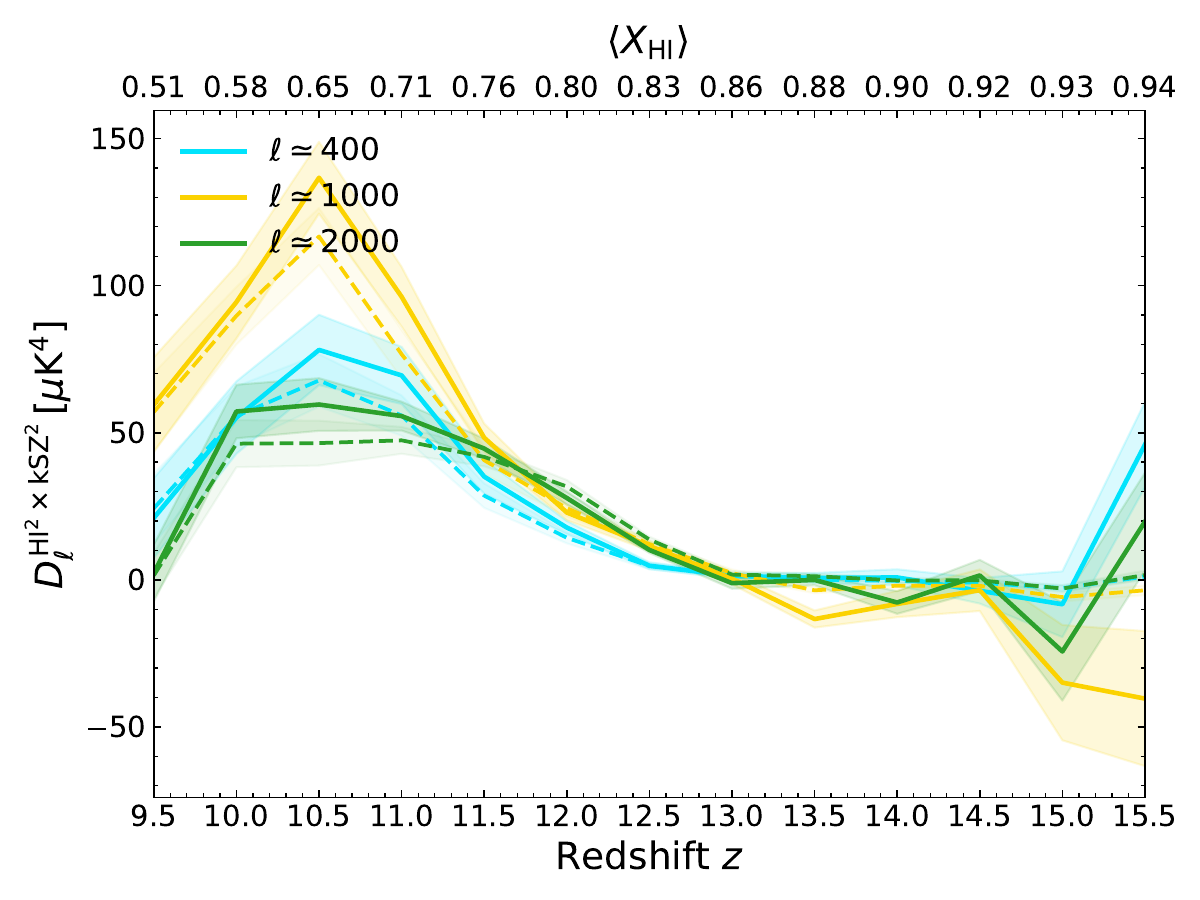}
    \caption{ {Redshift evolution of the $\mathrm{kSZ}^{2}\times21\mathrm{cm}^{2}$ power spectrum at representative multipoles in the early-heating model. Solid curves include spatial fluctuations in the HI spin temperature, while dashed curves adopt the saturated-heating approximation $T_{\rm s}\gg T_{\rm CMB}$. The 21-cm field is evaluated in redshift chunks of width $\Delta z=1$, with foreground avoidance excluding modes with $k_{\parallel,0}=0.1\,h\,{\rm Mpc}^{-1}$ and instrumental-noise filtering for the SKA-Low AA$^\ast$ configuration. The kSZ field is filtered using the CMB-HD configuration. In contrast to the fiducial model (Fig.~\ref{fig:Redshift_Evolution_Ts_vs_noTs}), the two cases agree closely across most of the redshift range.}}
    \label{fig:Redshift_Evolution_Ts_vs_noTs_early}
\end{figure}

\subsection{Dependence on Heating History}
To obtain an early-heating history, we vary several key parameters; specifically, we increase $\log_{10}(F_{\ast,10})$ from $-1.25$ to $-1.20$, corresponding to a modest enhancement in the star-formation efficiency. We further increase the X-ray luminosity per unit star-formation rate, $\log_{10}(L_{\rm X}/{\rm SFR})$, from $40.5$ to $41.52$, enhancing the X-ray emissivity by approximately an order of magnitude. Finally, we lower the X-ray absorption threshold $E_0$ from $500$ to $300\,{\rm eV}$, allowing a larger fraction of lower-energy X-ray photons to escape their host galaxies and contribute to heating the IGM. Together, these modifications lead to more efficient and earlier heating of the IGM, resulting in an earlier evolution of the spin temperature during cosmic dawn and the early stages of reionization. The resulting scenario therefore provides a useful test of the sensitivity of the $\mathrm{kSZ}^2\times21\mathrm{cm}^2$ cross-correlation to different thermal histories of the IGM.

We compare the cross-correlation signals in the fiducial and early-heating scenarios in Fig.~\ref{fig:DifferentReionization_history}. We consider redshift chunks of width $\Delta z=1$ and adopt CMB-HD and SKA-Low AA$^\ast$ for the kSZ and 21-cm observations, respectively. An optimistic foreground-avoidance filter is applied to the 21-cm field, excluding modes with $k_\parallel\leq 0.1\,h\,{\rm Mpc}^{-1}$. The left panel shows the cross-correlation signal as a function of the volume-averaged neutral fraction, $\langle {x}_{\rm HI}\rangle$, while the right panel shows the same signal as a function of the ratio $\langle T_{\rm CMB}/T_{\rm s}\rangle$. In both panels, the solid and dashed curves represent the fiducial and early-heating scenarios, respectively.

\begin{figure}
    \centering
    \includegraphics[width=1.\linewidth]{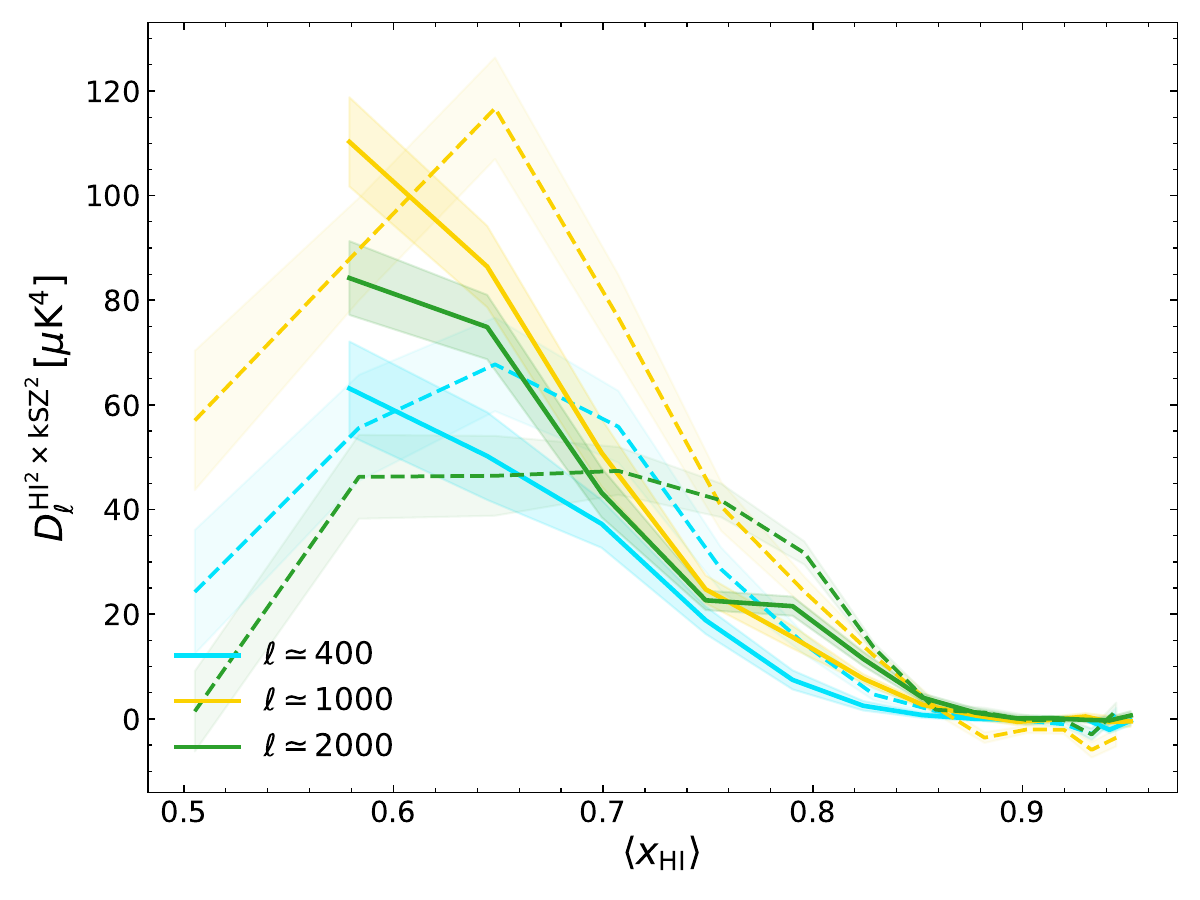}
    \caption{ {The $\mathrm{kSZ}^{2}\times21\mathrm{cm}^{2}$ cross-correlation under the saturated-heating approximation, in which $1-T_{\rm CMB}/T_{\rm s}$ is set to unity, shown as a function of the volume-averaged neutral fraction $\langle x_{\rm HI}\rangle$ for the fiducial (solid) and early-heating (dashed) scenarios. In this limit the signal carries no information on the thermal history and is determined solely by the ionization field. The cross-correlation is evaluated in redshift chunks of width $\Delta z=1$, assuming CMB-HD and SKA-Low AA$^\ast$ for the kSZ and 21-cm observations, respectively. An optimistic foreground-avoidance filter with $k_{\parallel,0}=0.1\,h\,{\rm Mpc}^{-1}$ is applied to the 21-cm field.}}
    \label{fig:DifferentReionization_history_xHI_noTs}
\end{figure}

Interestingly, when the results from the two scenarios are combined, the cross-correlation signal shows no clear dependence on the neutral fraction. In the fiducial model, a significant cross-correlation is obtained when the Universe is more neutral, around $\langle {x}_{\rm HI}\rangle \approx0.9$, whereas the signal in the early-heating scenario remains weak at a similar neutral fraction and becomes significant only when the Universe is more considerably ionized, at $\langle {x}_{\rm HI}\rangle \sim0.7$. This indicates that the neutral fraction alone is insufficient to characterize the evolution of the $\mathrm{kSZ}^2\times21\mathrm{cm}^2$ signal.

\begin{figure*}
    \centering
    \includegraphics[width=0.48\linewidth]{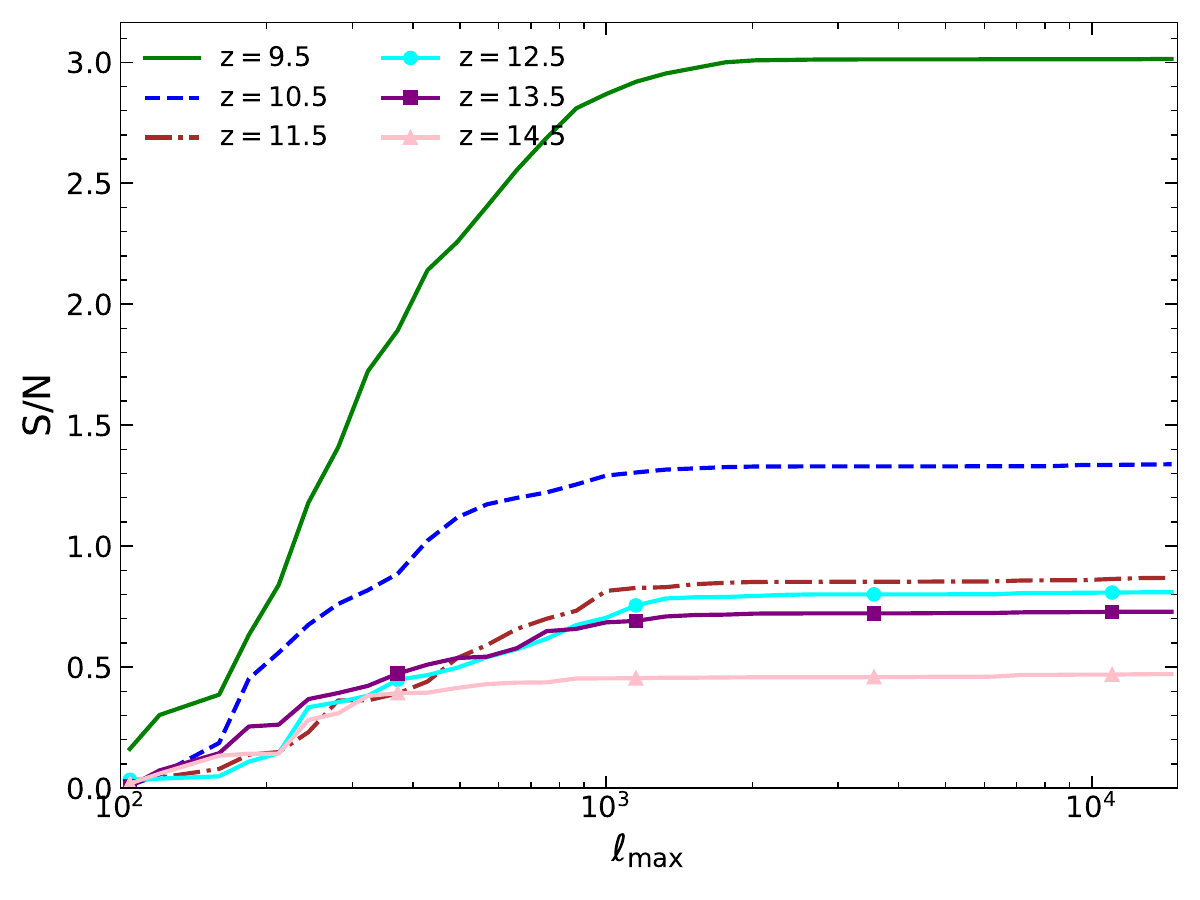}
    \includegraphics[width=0.48\linewidth]{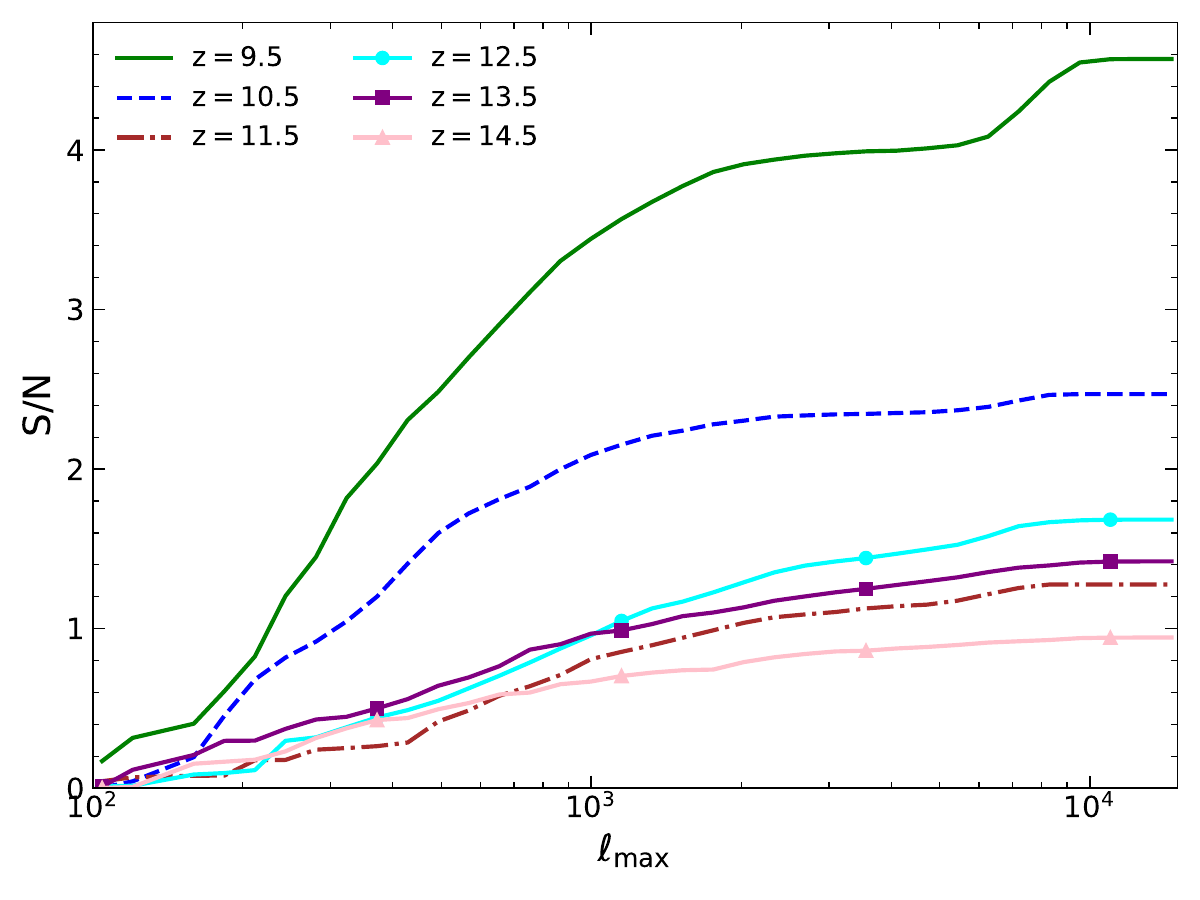}
    \caption{Signal-to-noise ratio of the $\mathrm{kSZ}^{2}\times21\,\mathrm{cm}^{2}$ cross-correlation in individual redshift bins for the fiducial model, cumulatively evaluated up to $\ell_{\max}$ by summing over multipoles, with a chunk size of $\Delta z=1$. \textit{Left}--HERA$\times$CMB-HD. \textit{Right}--SKA-Low AA$^\ast\times$CMB-HD. In both cases, modes with $k_{\parallel}\leq0.1\,h\,{\rm Mpc}^{-1}$ are excluded from the 21-cm field.}
    \label{fig:SNRz_fiducial}
\end{figure*}

In contrast, a much clearer dependence on the spin temperature is found
in the right panel. In both scenarios, the cross-correlation reaches a
minimum when the spin temperature approaches the CMB temperature,
corresponding to
\begin{equation}
\left\langle 1-\frac{T_{\rm CMB}}{T_{\rm s}}\right\rangle \simeq 0.
\end{equation}
On either side of this minimum, the cross-correlation signal in both scenarios follows a qualitatively similar evolution with $\langle T_{\rm CMB}/T_{\rm s}\rangle$, indicating that the signal is strongly modulated by the HI spin temperature. Nevertheless, the two scenarios peak on opposite sides of the minimum: the fiducial model peaks at $\langle T_{\rm CMB}/T_{\rm s}\rangle \simeq 2.5$, whereas the early-heating model peaks at $\langle T_{\rm CMB}/T_{\rm s}\rangle \simeq 0.1$. Both peaks correspond to positive correlations, but they arise from different physical conditions.

The key physical difference is whether the 21-cm signal is in emission or absorption at the peak epoch. Because the 21-cm field enters the statistic as a square, the overall sign of the 21-cm signal is lost, and the cross-correlation alone cannot directly distinguish between the two cases. However, the side of the minimum on which the peak occurs still carries this information. In the fiducial model, the peak lies on the $T_{\rm s} < T_{\rm CMB}$ side, so that the 21-cm signal is in absorption. In the early-heating model, the peak lies on the $T_{\rm s} \gg T_{\rm CMB}$ side, so that the 21-cm signal is in emission.

To illustrate this difference, we show slices of the 21-cm signal at the epoch of the peak cross-correlation for both scenarios in Fig.~\ref{fig:HI_cmbz_Ts}. In the fiducial case (left panel), the region surrounding the ionized bubbles is predominantly in absorption. In the early-heating case (right panel), almost the entire IGM exhibits 21-cm emission, with only small $\delta T_{\rm b} = 0$ regions corresponding to fully ionized bubbles. Thus, while the squared 21-cm field erases the sign of the signal, the position of the cross-correlation peak relative to $\langle T_{\rm CMB}/T_{\rm s}\rangle = 1$ still provides an indirect diagnostic of whether the 21-cm signal is in emission or absorption, and the slices in Fig.~\ref{fig:HI_cmbz_Ts} directly confirm this interpretation.

 {To quantify the regime in which spin-temperature fluctuations remain important, we repeat the comparison of Sec.~\ref{subsec:SpinTemperature} for the early-heating scenario, computing the cross-correlation both with the full 21-cm signal and under the saturated-heating approximation in which $1 - T_{\rm CMB}/T_{\rm s}$ is set to unity. The two calculations are shown as 
solid and dashed lines in Fig.~\ref{fig:Redshift_Evolution_Ts_vs_noTs_early}, respectively. As before, we consider redshift chunks of width $\Delta z=1$, 
adopt CMB-HD and SKA-Low AA$^\ast$ for the kSZ and 21-cm observations, and exclude modes with $k_{\parallel} \leq 0.1\,h\,{\rm Mpc}^{-1}$ from the 21-cm field. In contrast to the fiducial model, where the two cases differ substantially at $z \gtrsim 12$ (Fig.~\ref{fig:Redshift_Evolution_Ts_vs_noTs}), the early-heating results agree to within the $2\sigma$ uncertainties across essentially the entire redshift range considered, with small deviations appearing only at $z \gtrsim 14$, where $\langle T_{\rm s}\rangle$ has not yet risen considerably above $T_{\rm CMB}$.}

 {The absence of a boost from $T_{\rm s}$ fluctuations in the early-heating scenario can be understood as follows. Heating and ionization both commence around Cosmic Dawn, but their relative timing is set by the X-ray heating efficiency. When heating is sufficiently efficient, the IGM absorbs enough X-ray photons to raise $T_{\rm s}$ well above $T_{\rm CMB}$ before reionization has advanced appreciably. Spatial fluctuations in $T_{\rm s}$ are largely confined to the earliest epochs ($z \gtrsim 14$ in this model), where the IGM remains almost fully neutral ($\langle x_{\rm HI}\rangle \gtrsim 0.9$) and the patchy kSZ signal is correspondingly weak. There is thus little kSZ power for these fluctuations to correlate with, and the enhancement seen in the fiducial model never develops; by the time ionized regions have grown large enough to source an appreciable kSZ signal, $T_{\rm s}$ has nearly saturated. The sensitivity of the $\rm kSZ^2 \times 21\,cm^2$ cross-correlation to the thermal history is therefore governed by the efficiency of X-ray heating: it is pronounced in the fiducial model, where the absorption epoch overlaps with the growth of ionized structure, and becomes weak in the early-heating model, in which the enhanced X-ray emissivity and softer absorption threshold drive heating to completion before reionization is well underway.}

 {We note that the early-heating scenario is obtained by varying $\log_{10}(F_{*,10})$ in addition to the X-ray parameters relative to the fiducial model; this parameter affects not only the heating history but also the progression of reionization. To assess how much the modified reionization history contributes to the differences between the two scenarios, we compare the cross-correlation signals computed for the fiducial and early-heating models, each under the saturated-heating approximation, shown as a function of $\langle x_{\rm HI}\rangle$ in Fig.~\ref{fig:DifferentReionization_history_xHI_noTs}. In this limit the signal carries no information on the thermal history and is determined solely by the ionization field, essentially the configuration considered by \citet{Zhou:2025fgv}; plotting against $\langle x_{\rm HI}\rangle$ rather than redshift further removes any difference in the timing of reionization, so that only the ionization morphology can differ between the two cases. The solid and dashed curves, corresponding to the fiducial and early-heating models respectively, are broadly consistent in both shape and amplitude. By contrast, the full calculations shown in the left panel of Fig.~\ref{fig:DifferentReionization_history} differ substantially, with the two models peaking at markedly different neutral fractions. We therefore conclude that the differences between the two scenarios are driven primarily by their distinct thermal histories, with the modest change in the ionization history playing only a minor role.}

\begin{figure*}[t]
    \centering
    \includegraphics[width=0.48\linewidth]{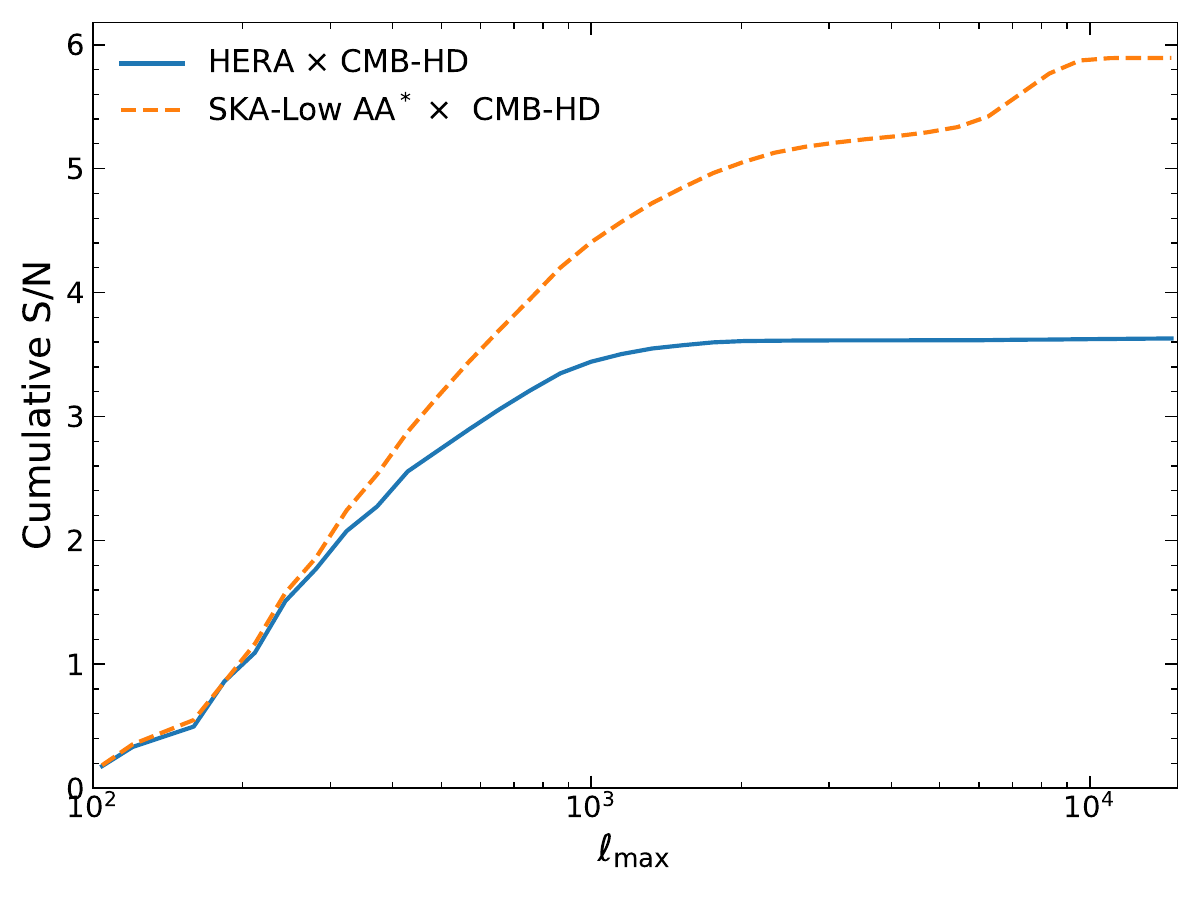}
    \includegraphics[width=0.48\linewidth]{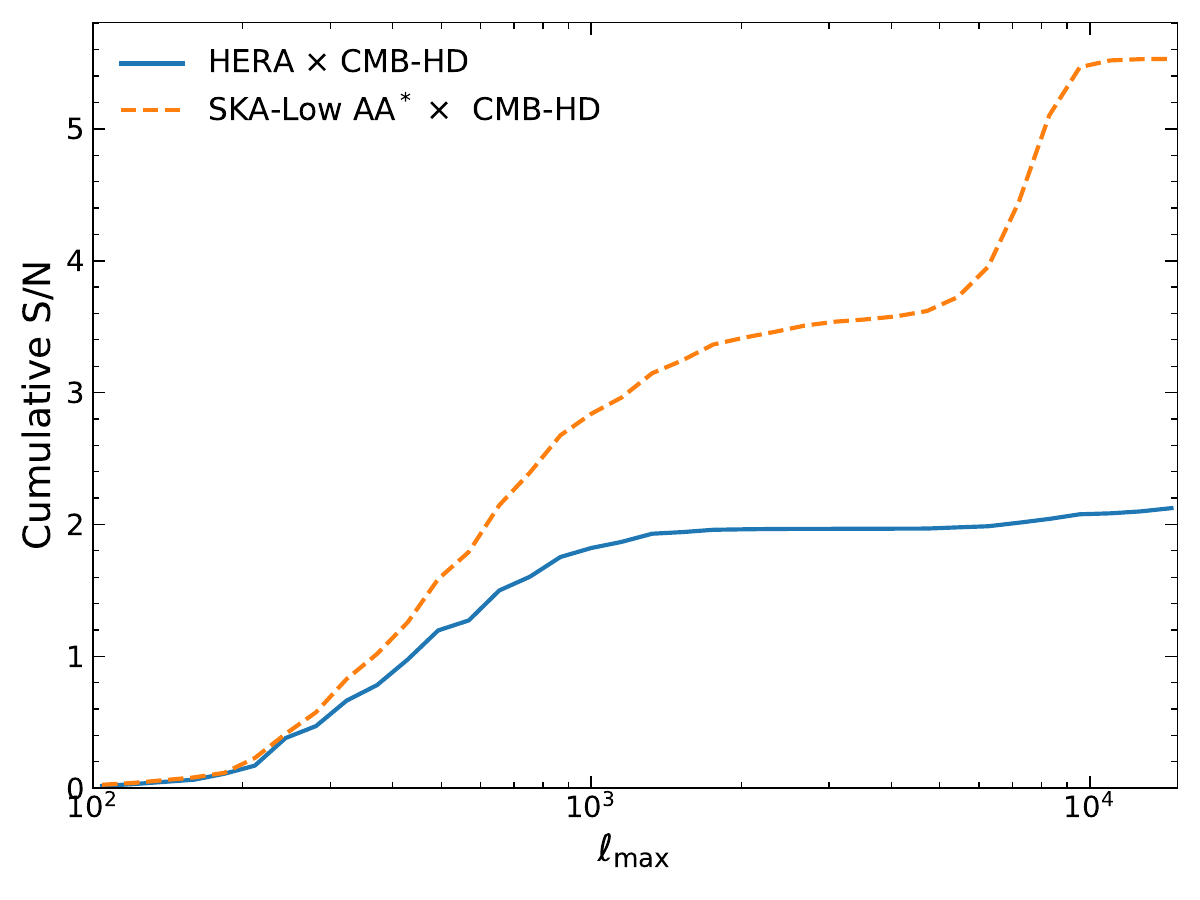}
    \caption{Cumulative signal-to-noise ratio of the $\mathrm{kSZ}^{2}\times21\,\mathrm{cm}^{2}$ cross-correlation as a function of $\ell_{\max}$. \textit{Left}--Fiducial model. \textit{Right}--Early-heating scenario. The blue and orange curves correspond to CMB-HD combined with HERA and SKA-Low AA$^\ast$, respectively. A redshift chunk width of $\Delta z=1$ is adopted, with modes satisfying $k_\parallel\leq0.1\,h\,{\rm Mpc}^{-1}$ excluded from the 21-cm field.
}
    \label{fig:CumulativeSNR}
\end{figure*}

\section{Detectability Forecast and Discussion}
\label{sec:detectability}

In this section, we estimate the detectability of the $\mathrm{kSZ}^2\times21\,\mathrm{cm}^2$ cross-correlation using current and upcoming surveys. As discussed above, we consider the cross-correlation between the 21-cm signal measured by HERA or SKA-Low AA$^\ast$ and the kSZ signal observed by the SO or CMB-HD. We first estimate the signal-to-noise ratio for a given experimental combination and redshift bin centered at $z_0$, summing over all available angular multipoles $\ell$. The corresponding S/N can be written as \citep{Ferraro:2016ymw}
\begin{equation}
\left(\frac{S}{N}\right)_{z_{0}}^{2}=\sum^{\ell_{\rm max}}_{\ell=100}\frac{f_{\mathrm{sky}}(2\ell+1)\left(C_{\ell}^{\mathrm{kSZ}^{2}\times 21\,\mathrm{cm}^{2}}\right)^{2}}{C_{\ell}^{T^{2}\!}C_{\ell}^{(21\,\mathrm{cm}+{\rm N})^{2}}+\left(C_{\ell}^{\mathrm{kSZ}^{2}\times 21\,\mathrm{cm}^{2}}\right)^{2}},
\label{eq:snr_single}
\end{equation}
where $f_{\mathrm{sky}}$ is the fraction of sky jointly covered by the two experiments. In this work, we adopt a fiducial overlapping survey area of $2000\,{\rm deg}^{2}$, corresponding to $f_{\rm sky}=0.048$, motivated by the approximate overlap between the SO large-aperture survey and the HERA observing footprint. For SKA-Low AA$^\ast$, the accessible sky area is expected to be larger, so this choice is conservative for SKA-based combinations. Here, $C_{\ell}^{T^{2}}$ denotes the auto-power spectrum of the CMB temperature map after Fourier-space filtering and subsequent squaring.

We follow the approach of \cite{Ferraro:2016ymw} and \cite{LaPlante:2021ced} to estimate $C_{\ell}^{T^{2}}$ under the Gaussian approximation
\begin{equation}
C_{\ell}^{T^{2}}\approx 2\int \frac{{\rm d}^{2}\boldsymbol{L}}{(2\pi)^{2}}C_{L}^{T}C_{|\boldsymbol{L}-\boldsymbol{\ell}|}^{T},
\label{eq:cl_t2t2f}
\end{equation}
where
\begin{equation}
C_{\ell}^{T}=f^{\mathrm{kSZ}}(\ell)^{2}\left(C_{\ell}^{TT}+C_{\ell}^{\mathrm{kSZ}}|_{\rm EoR}+C_{\ell}^{\mathrm{kSZ}}|_{\rm late}+N_{\ell}\right).
\label{eq:cl_ttf}
\end{equation}

In contrast, $C_{\ell}^{(21\,\mathrm{cm}+{\rm N})^{2}}$ denotes the auto angular power spectrum obtained from the 21-cm field with instrumental noise included. Since the 21-cm field during the EoR is highly non-Gaussian, we evaluate this power spectrum directly from simulations by generating Gaussian noise realizations in Fourier space from the noise power spectrum and adding them to the simulated 21-cm signal, then applying the same filtering, squaring, and line-of-sight projection procedures as used in constructing the $\mathrm{kSZ}^{2}\times21\,\mathrm{cm}^{2}$ cross-correlation.

The S/N in each redshift bin with a chunk width of $\Delta z=1$ for our fiducial model is shown in Fig.~\ref{fig:SNRz_fiducial}. The left and right panels correspond to the cases in which the 21-cm signal is observed by HERA and SKA-Low AA$^\ast$, respectively, while the kSZ signal is observed by CMB-HD. In both cases, we exclude modes satisfying $k_\parallel\leq 0.1\,h\,{\rm Mpc}^{-1}$. The highest S/N is obtained at $z\sim 9$, where the impact of the spin temperature on the cross-correlation is relatively weak.

At higher redshifts, corresponding to the regime in which spin-temperature evolution becomes increasingly important ($z\gtrsim11$), the cross-correlation signal can remain pronounced, as shown in Fig.~\ref{fig:Redshift_Evolution_Ts_vs_noTs}. However, the auto-power spectrum $C_{\ell}^{(21\,\mathrm{cm}+{\rm N})^{2}}$ also becomes significant in this regime, leading to an enhanced variance and hence limiting the resulting S/N. We find that this contribution is dominated by the intrinsic fluctuations of the 21-cm field rather than by instrumental noise, since the latter is efficiently suppressed by the Wiener filtering adopted in our analysis. Therefore, although spin-temperature fluctuations can significantly modify the amplitude and redshift evolution of the cross-correlation signal, this does not necessarily translate into a high detection significance.

The difference in S/N between HERA and SKA-Low AA$^\ast$ can also be understood from their sensitivity to different transverse Fourier modes. As discussed above, HERA primarily probes relatively large transverse scales, while the more extended baseline distribution of SKA-Low AA$^\ast$ provides sensitivity to intermediate and smaller scales. Consequently, the S/N for HERA saturates at $\ell\sim10^{3}$, whereas that for SKA-Low AA$^\ast$ continues to increase to higher multipoles and saturates at approximately $\ell\sim10^{4}$. This behavior reflects the different angular-scale coverage of the two 21-cm experiments and further demonstrates that the detectability of the $\mathrm{kSZ}^{2}\times21\,\mathrm{cm}^{2}$ signal depends not only on its intrinsic amplitude, but also on the range of transverse modes accessible to the experiment. It is worth noting that SKA-Low additionally supports substation configurations, where a single station can be subdivided to access shorter baselines. Including such a configuration would improve the sensitivity of SKA-Low AA$^\ast$ at low multipoles, potentially bridging the gap with HERA and further strengthening the complementarity between the two arrays for measuring the $\mathrm{kSZ}^2\times21\,\mathrm{cm}^2$ cross-correlation.

\begin{deluxetable}{lcccc}
\tablecaption{Cumulative signal-to-noise ratios of the $\mathrm{kSZ}^{2}\times21\,\mathrm{cm}^{2}$ cross-correlation for HERA combined with different CMB experiments. A redshift chunk width of $\Delta z=1$ is adopted. The columns correspond to different foreground-avoidance prescriptions for the 21-cm field: excluding modes with $k_\parallel\leq 0.01\,h\,{\rm Mpc}^{-1}$, excluding modes with $k_\parallel\leq0.10\,h\,{\rm Mpc}^{-1}$, and applying the redshift-dependent wedge cut $k_\parallel\leq m(z)k_\perp$.}
\label{tab:cumulative_snr_hera}
\tablehead{
\colhead{CMB Experiment} &
\colhead{Reionization Scenario} &
\colhead{$k_{\parallel,0}=0.01$} &
\colhead{$k_{\parallel,0}=0.10$} &
\colhead{$m(z)$}
}
\startdata
SO     & Fiducial      & 0.64 & 0.43 & 0.31 \\
       & Early-heating & 0.27 & 0.22 & 0.21 \\
CMB-HD & Fiducial      & 5.46 & 3.62\tablenotemark{a} & 2.85 \\
       & Early-heating & 2.57 & 2.12 & 1.60
\enddata
\tablenotetext{a}{ {In the noise-limited case where foreground residuals in the CMB map are fully removed, with $N_{\ell} = \Delta_{\rm N}^{2}b^{-2}(\ell)$ and $\Delta_{\rm N} = 0.6\,{\rm \mu K\text{-}arcmin}$ for CMB-HD \citep{LaPlante:2021ced}, the cumulative S/N increases to $9.00$.}}
\end{deluxetable}

We can extend the S/N calculation for a single redshift bin to determine the cumulative S/N over the full redshift range. The cumulative S/N can then be written as
\begin{equation}
\left(S/N\right)_{\mathrm{cum}} = \sqrt{\sum^{16}_{z_{0}=9} (S/N)_{z_{0}}^{2}}.
\label{eq:snr_tot}
\end{equation}

In Fig.~\ref{fig:CumulativeSNR}, we show the cumulative S/N for the two reionization scenarios, with the fiducial model presented in the left panel and the early-heating scenario in the right panel, assuming a redshift chunk size of $\Delta z = 1$. In both panels, the blue solid curve represents the cumulative S/N for HERA combined with CMB-HD, while the orange curve corresponds to SKA-Low AA$^\ast$ combined with CMB-HD. In all cases, we adopt the optimistic foreground-avoidance criterion and exclude modes satisfying $k_\parallel\leq 0.1\,h\,{\rm Mpc}^{-1}$.

Finally, we summarize the cumulative S/N for the different combinations of CMB and 21-cm experiments and foreground-removal strategies in Tables~\ref{tab:cumulative_snr_hera} and \ref{tab:cumulative_snr_SKA_AAstar}, assuming a redshift chunk width of $\Delta z=1$. Table~\ref{tab:cumulative_snr_hera} presents the results for HERA combined with SO and CMB-HD, while Table~\ref{tab:cumulative_snr_SKA_AAstar} shows the corresponding results for SKA-Low AA$^\ast$. In each case, we consider three foreground-avoidance scenarios spanning progressively more conservative assumptions. We first adopt an optimistic, nearly idealized foreground-removal scenario with $k_{\parallel,0}=0.01\,h\,{\rm Mpc}^{-1}$, in which foreground contamination is assumed to be confined to a narrow region around the lowest line-of-sight modes, allowing most of the $k_\parallel$ modes to be retained. We then consider a more conservative scenario with $k_{\parallel,0}=0.10\,h\,{\rm Mpc}^{-1}$, corresponding to a larger loss of low-$k_\parallel$ modes due to residual foreground contamination and imperfect foreground removal. Finally, we adopt the redshift-dependent foreground wedge criterion, $k_\parallel \leq m(z)k_\perp$, which accounts for the coupling between transverse and line-of-sight modes induced by the chromatic response of the interferometer and provides a more realistic estimate of the impact of foreground avoidance.

\begin{deluxetable}{lcccc}
\tablecaption{Cumulative signal-to-noise ratios of the $\mathrm{kSZ}^{2}\times21\,\mathrm{cm}^{2}$ cross-correlation for SKA-Low AA$^\ast$ combined with different CMB experiments. A redshift chunk width of $\Delta z=1$ is adopted. The columns correspond to different foreground-avoidance prescriptions for the 21-cm field: excluding modes with $k_\parallel\leq 0.01\,h\,{\rm Mpc}^{-1}$, excluding modes with $k_\parallel\leq 0.10\,h\,{\rm Mpc}^{-1}$, and applying the redshift-dependent wedge cut $k_\parallel \leq m(z)k_\perp$.
\label{tab:cumulative_snr_SKA_AAstar}}
\tablehead{
\colhead{CMB Experiment} &
\colhead{Reionization Scenario} &
\colhead{$k_{\parallel,0}=0.01$} &
\colhead{$k_{\parallel,0}=0.10$} &
\colhead{$m(z)$}
}
\startdata
SO     & Fiducial      & 1.40 & 0.90 & 0.34 \\
       & Early-heating & 1.61 & 1.08 & 0.21 \\
CMB-HD & Fiducial      & 8.20 & 5.89\tablenotemark{b} & 2.94 \\
       & Early-heating & 7.44 & 5.53 & 2.02
\enddata
\tablenotetext{b}{ {Increases to $13.37$ in the noise-limited case; see note $^a$ in Table~\ref{tab:cumulative_snr_hera}.}}
\end{deluxetable}

Several trends are evident from these results. First, the detectability is more strongly affected by the choice of CMB experiment than by the choice of 21-cm experiment for the configurations considered here. For example, in the fiducial model, under the $k_{\parallel,0}=0.10\,h\,{\rm Mpc}^{-1}$ foreground cut, the cumulative S/N increases from $0.43$ to $3.62$ for HERA and from $0.90$ to $5.89$ for SKA-Low AA$^\ast$ when replacing SO with CMB-HD. This improvement reflects the importance of the CMB instrumental noise and angular resolution in extracting the kSZ contribution and constructing the filtered squared CMB field. In contrast, changing the 21-cm experiment from HERA to SKA-Low AA$^\ast$ produces a more moderate change in the cumulative S/N. Given this sensitivity to the CMB experiment, we perform an additional calculation to quantify the prospects for a noise-limited CMB measurement. We consider an idealized case in which the residual foregrounds in the CMB map,  including the cosmic infrared background (CIB), radio sources, and the thermal Sunyaev--Zel'dovich (tSZ) effect, are fully removed, so that $N_{\ell}$ consists solely of instrumental noise. For CMB-HD, this pure noise term can be written as $N_{\ell} = \Delta_{\rm N}^{2}b^{-2}(\ell)$ with $\Delta_{\rm N} = 0.6\,{\rm \mu K\text{-}arcmin}$ \citep{LaPlante:2021ced}. In this noise-limited limit, the cumulative S/N further increases to $9.00$ and $13.37$ for CMB-HD combined with HERA and SKA-Low AA$^\ast$, respectively {, and are reported as footnotes in 
Tables~\ref{tab:cumulative_snr_hera} and~\ref{tab:cumulative_snr_SKA_AAstar}. These values represent optimistic upper bounds on the achievable S/N for a noise-limited CMB-HD measurement.}

Second, the detectability depends appreciably on the foreground-removal
strategy. Increasing the minimum retained line-of-sight wavenumber from $k_{\parallel,0}=0.01$ to $0.10\,h\,{\rm Mpc}^{-1}$ reduces the cumulative S/N for both 21-cm experiments and both reionization scenarios. The suppression becomes particularly severe when the redshift-dependent wedge criterion is adopted. This demonstrates that the loss of low-$k_\parallel$ modes can limit the detectability of the $\mathrm{kSZ}^{2}\times21\,\mathrm{cm}^{2}$ cross-correlation. It is important to reiterate, however, that this statistic behaves differently from simpler cross-correlations: the latter vanish under foreground filtering, whereas a non vanishing signal survives here. Foreground filtering limits the amplitude of the signal rather than erasing it entirely, which highlights the advantage of this higher-order statistic over simpler cross-correlations.

 {Third, we note that switching from the fiducial to the early-heating 
scenario has a markedly different impact on the cumulative S/N for HERA 
and SKA-Low AA$^\ast$. For SKA-Low AA$^\ast$, the cumulative S/N is comparable between the fiducial and early-heating scenarios (Table~\ref{tab:cumulative_snr_SKA_AAstar}), whereas 
for HERA the early-heating scenario yields a suppression by a factor of 
$\sim 2$ relative to the fiducial model across all foreground cases 
(Table~\ref{tab:cumulative_snr_hera}). This difference can be understood from 
the interplay between the thermal history and the ionization structure, and their respective contributions to the cross-correlation signal at different angular scales.}

\begin{figure}
    \centering
    \includegraphics[width=1.\linewidth]{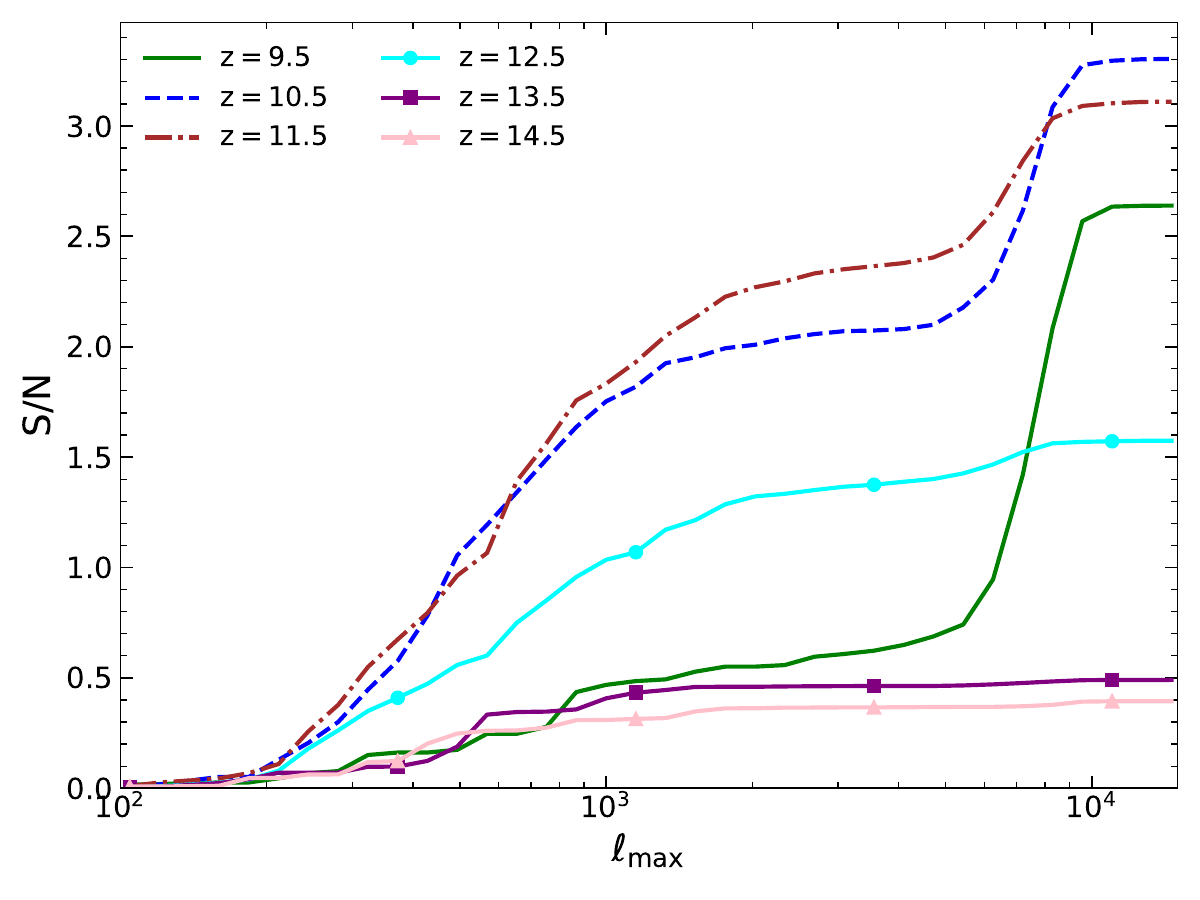}
    \caption{ {Signal-to-noise ratio of the $\mathrm{kSZ}^{2}\times21\,\mathrm{cm}^{2}$ 
    cross-correlation in individual redshift bins for the early-heating scenario, 
    cumulatively evaluated up to $\ell_{\rm max}$ by summing over multipoles, 
    assuming SKA-Low AA$^\ast\times$CMB-HD with a chunk size of $\Delta z=1$. 
    Modes with $k_{\parallel}\leq0.1\,h\,{\rm Mpc}^{-1}$ are excluded from the 
    21-cm field. This figure complements the right panel of 
    Fig.~\ref{fig:SNRz_fiducial} and is provided to aid the interpretation of 
    the difference in cumulative S/N between the fiducial and early-heating 
    scenarios reported in Table~\ref{tab:cumulative_snr_hera} and 
    Table~\ref{tab:cumulative_snr_SKA_AAstar}.}}
    \label{fig:SNRz_SKA_CMBHD_early}
\end{figure}

 {
To illustrate this, we show in Fig.~\ref{fig:SNRz_SKA_CMBHD_early} the S/N as a function of $\ell_{\rm max}$ for individual redshift bins in the early-heating scenario, assuming SKA-Low AA$^\ast \, \times$ CMB-HD with $k_{\parallel,0} = 0.1\,h\,{\rm Mpc}^{-1}$. At $z \gtrsim 12$, the S/N accumulates gradually across all scales, with no pronounced small-scale contribution. However, at $z \lesssim 12$, clear upturns in the S/N emerge at small angular scales ($\ell \gtrsim 6000$), indicating that small-scale modes begin to contribute significantly to the signal. As shown in the lower panel of Fig.~\ref{fig:NeutralFraction_BrightnessTemperature}, this transition 
coincides with the epoch at which $\langle T_{\rm s} \rangle$ begins to 
exceed $T_{\rm CMB}$ in the early-heating scenario. As $T_{\rm s}$ rises above $T_{\rm CMB}$, spatial fluctuations in $T_{\rm s}$ progressively lose their ability to modulate $\delta T_{\rm b}$, and the $\rm kSZ^2 \times 21\,cm^2$ signal becomes increasingly dominated by fluctuations in the ionization field, with small ionized bubbles contributing prominently at small angular scales.}

 {
The same behaviour is observed in the fiducial model, as seen in the right panel of Fig.~\ref{fig:SNRz_fiducial}: a small-scale upturn in the S/N appears only at $z \lesssim 10$, again coinciding with the epoch at which $\langle T_{\rm s} \rangle$ exceeds $T_{\rm CMB}$ in the fiducial model (lower panel of Fig.~\ref{fig:NeutralFraction_BrightnessTemperature}). This consistency across both scenarios confirms that the emergence of small-scale contributions to the cross-correlation signal is governed by the epoch at which $T_{\rm s}$ fluctuations become subdominant, after which the ionization structure takes over as the primary driver of the signal.}

 {
Consequently, within $z = 9$--$16$, the early-heating scenario contains small-scale contributions over a broader redshift range ($z \sim 9$--$12$) than the fiducial model ($z \lesssim 10$). HERA, being primarily sensitive to large-to-intermediate scales, misses a larger fraction of this small-scale signal in the early-heating scenario, leading to the observed suppression of the cumulative S/N. SKA-Low AA$^\ast$, by contrast, retains sensitivity up to $\ell \sim 10^4$ and captures the small-scale signal in both scenarios, resulting in comparable cumulative S/N values across the two heating histories.
}

Throughout this work, we adopt a redshift chunk width of $\Delta z=1$ for the calculation of the $\mathrm{kSZ}^{2}\times21\,\mathrm{cm}^{2}$ cross-correlation and its signal-to-noise ratio. To test the robustness of our results, we also examined choices of $\Delta z = 0.5$ and $\Delta z = 2$, finding that the cumulative S/N varies only moderately with the chunk size. A rigorous optimization of the redshift binning, which would require balancing the redshift resolution of the reionization history, foreground leakage into high-$k_\parallel$ modes, and inter-bin covariances for parameter estimation, is beyond the scope of this work and is left for future investigation.

\section{Conclusion}\label{sec:conclusion}
We have investigated the $\mathrm{kSZ}^{2}\times21\,\mathrm{cm}^{2}$ angular cross-correlation as a probe of the early stages of the epoch of reionization (EoR), using angular lightcones generated with the semi-numeric simulation \texttt{21cmFASTv4}. Unlike the linear $\mathrm{kSZ}\times21\,\mathrm{cm}$ cross-correlation, which is strongly suppressed by the cancellation of positive and negative kSZ fluctuations, and the cross-correlation of a singly squared kSZ field with HI, $\mathrm{kSZ}^{2}\times21\,\mathrm{cm}$, in which almost all $k_{\parallel}$-mode information is lost due to foreground contamination, the statistic considered here involves the squared fields of both the kSZ and 21-cm signals. It therefore retains information from pairs of opposite, nonzero $k_\parallel$ modes, making the $\mathrm{kSZ}^{2}\times21\,\mathrm{cm}^{2}$ cross-correlation particularly suitable for extracting EoR information from modes that survive the foreground wedge.

We find that the cross-correlation is sensitive not only to the ionization morphology but also to the thermal state of the neutral IGM. In particular, spatial fluctuations in the HI spin temperature can modify both the amplitude and the redshift evolution of the signal.  {The sensitivity to the thermal state, however, depends on the efficiency of X-ray heating: when heating is efficient enough that the IGM reaches $T_{\rm s}\gg T_{\rm CMB}$ before reionization is well underway, as in our early-heating scenario, spin-temperature fluctuations are confined to epochs where the patchy kSZ signal is still weak, and the saturated-heating approximation becomes adequate.} Consequently,  {provided that heating is not too efficient,} the redshift evolution of the $\mathrm{kSZ}^{2}\times21\,\mathrm{cm}^{2}$ signal contains information about the thermal history of the IGM that is complementary to constraints based solely on the global neutral fraction.

We have also assessed the detectability of the signal for combinations of HERA or SKA-Low AA$^\ast$ with the Simons Observatory and CMB-HD. The cumulative signal-to-noise ratio is generally more sensitive to the CMB experiment than to the choice of 21-cm experiment for the configurations considered here. Current CMB sensitivity, represented by the Simons Observatory, is insufficient for a significant detection over $z=9$--$16$, whereas the lower noise and finer angular resolution of CMB-HD can greatly improve the detectability. However, foreground on the 21-cm side remains a critical limitation. Removing progressively larger regions of low-$k_\parallel$ modes reduces the cumulative signal-to-noise ratio, and a wedge-based foreground criterion can largely diminish the advantage provided by extended 21-cm baselines.

Overall, our results indicate that the $\mathrm{kSZ}^{2}\times21\,\mathrm{cm}^{2}$ cross-correlation provides a promising complementary probe of the early EoR, with sensitivity to  {the ionization structure of the IGM and, for sufficiently moderate X-ray heating, to its thermal evolution as well}. An important direction for future work is to go beyond the deliberately simple estimator adopted here: a more elaborate cross-trispectrum estimator that sums over many quadrilateral configurations of modes could further improve the signal-to-noise ratio and thus enable a higher-sensitivity detection. Future measurements combining high-resolution, low-noise CMB observations with improved recovery of 21-cm Fourier modes could therefore provide valuable constraints on the thermal and reionization history of the high-$z$ Universe.

\section{Acknowledgments}
We thank the useful discussion with Dr. Paul La Plante, Dr. Meng Zhou, and Prof. Léon V. E. Koopmans. 
W.-Q. Guo and Y.-Z. Ma acknowledge South Africa Research Chair Initiative (SARChI; Grant No. RCCA250325306198) from South Africa's Department of Science, Technology and Innovation and National Research Foundation (NRF), and other NRF Grants with No.~150580, No.~CHN22111069370, No.~ERC250324306141, No. AUPP250310302114, and also South African Radio Astronomy Observatory (SARAO) for its financial assistant (\url{www.sarao.ac.za}). We acknowledge computational cluster resources at the Centre for High-Performance Computing, Cape Town, South Africa.

\bibliography{kSZ2_HI2_ref}{}
\bibliographystyle{aasjournalv7}

\end{document}